%% file: total_eclipse.tex
\documentclass{aa}  

\usepackage{graphicx}
\usepackage{txfonts}
\usepackage{hyperref}
\usepackage{lscape}
\usepackage{multirow}
\usepackage{xcolor}
\newcommand{\creme}{CR\'EME}
\newcommand{\phitot}{$\Delta\phi_{\rm tot}$}
\newcommand{\ispec}{\texttt{iSpec}}
\newcommand{\jktebop}{\texttt{JKTEBOP}}

\begin{document}

   \title{High-resolution spectroscopy of detached eclipsing binaries during total eclipses}


   \author{K. G. He{\l}miniak
          \inst{1}
          \and
          J. M. Olszewska \inst{2}
          \and
          M. Puciata-Mroczynska \inst{3}
          \and 
          T. Pawar \inst{1}
          }

   \institute{Nicolaus Copernicus Astronomical Center, Polish Academy of Sciences, ul. Rabia\'{n}ska 8, 87-100 Toru\'{n}, Poland\\
              \email{xysiek@ncac.torun.pl}
         \and
             Astronomical Observatory Institute, Faculty of Physics, Adam Mickiewicz University, ul. S{\l}oneczna 36, 60-286 Pozna\'{n}, Poland
         \and
             Faculty of Physics, University of Warsaw, ul. Pasteura 5, 02-093 Warsaw, Poland
             }

   \date{Received April ..., 2024; accepted ..., 2024}

 
  \abstract
   {We present results of high-resolution spectroscopic observations of detached eclipsing binaries (DEBs) with total eclipses, for which UVES spectra were obtained during the phase of totality. These observations serve as a key way to determine the age and initial metallicity of the systems, as well as to verify evolutionary phases of their components and distances.}
   {With the additional, independent information on effective temperature and metallicity of one of the components, we aimed at estimating the precise ages of the studied binaries, and show the usefulness of totality spectra. The second goal was to provide precise orbital and physical stellar parameters of the components of systems in question.}
   {Using the VLT/UVES we obtained high-resolution spectra of 11 DEBs during their total eclipse phase. Atmospheric parameters of then-visible (larger) components were obtained with \ispec. With additional spectroscopy from the Comprehensive Research with \'Echelles on the Most interesting Eclipsing binaries (\creme) project, public archives, and literature, we obtained radial velocity (RV) measurements, from which orbital parameters were calculated. Photometric time-series observations from TESS and ASAS were modelled with the \texttt{JKTEBOP} code, and, combined with RV-based results, allowed us to obtain physical parameters for nine double-lined systems from our sample. All the available data were used to constrain the ages with our own approach, utilising \texttt{MESA} isochrones. Reddening-free, isochrone-based distances were also estimated, and confronted with {\it Gaia} Data Release 3 (GDR3) results.}
   {We show that single spectroscopic observations taken during a total eclipse can break the age-metallicity degeneracy, and allow for precise determination of the age of a DEB. With high-quality spectroscopic and photometric data, we are able to reach 5-10\% level of uncertainty (e.g. $724^{+52}_{-24}$~Myr). Even for single-lined DEBs, where absolute masses are not possible to obtain, the spectroscopic analysis of one of the components allows one to put strong constraints on the properties of both stars. For some cases we noted inconsistencies between isochrone-based and GDR3 distances. For one binary, which could not be fitted with a single isochrone (RZ~Eri) we suggest a new explanation.}
   {}


   \maketitle
%

\section{Introduction}
Detached eclipsing binaries (DEBs), especially those that are also double-lined spectroscopic binaries (SB2s) are one of the most important sources of direct determination of stellar parameters \citep{torr10}. The precision and accuracy of direct (and absolute) mass and radius measurements is unrivalled, thanks to the advent of high-stability spectrographs and satellite-borne photometric observations. Stellar parameters of hundreds of well-studied DEBs, from our Galaxy and the Magellanic Clouds \citep{debcat,eker14,grac14}, constitute the foundation of a number of astrophysical research, such as stellar formation and evolution, testing performance of satellite observations, characterisation of exoplanet hosts, or cosmic distance scale. Stars with high quality measurements, spanning a vast range of various stellar parameters, such as masses, spectral types, metallicities or evolutionary stages, are used as benchmark stars -- calibrators to validate analysis methods, cross-calibrate surveys, or improve data products of a survey \citep{bla14,max23}.

The assessment of precise and accurate effective temperatures ($T_{\rm eff}$) and metallicities ([$M/H$]) in DEBs (and SB2s in general) from spectroscopic observations is actually more complicated than in the case of single stars. The light that is instantaneously recorded comes from two sources, which can be very different from each other. One can be treated as contamination for the other. For this reason many DEBs, that are otherwise well studied, lack the information about metallicity, and temperatures of their components are calculated from colour indices (if multi-band photometry is available) and colour-temperature relations. For systems studied in the XX century the relations in question are now obsolete, but no modern attempts have been made to revisit those systems. Additionally, without [$M/H$] it is difficult to assess the age of a system, due to the age-metallicity degeneration. For example: on the main sequence (MS) a more metal-rich star is older than a metal-poor star of the same mass and $T_{\rm eff}$; on the red giant branch (RGB) isochrone fitting can give an error of $\sim$20~Myr for a fixed metallicity, while the error increases to $\sim$300~Myr while $\pm$0.1~dex uncertainty in [$M/H$] is introduced; etc. It is thus clear that for purposes of the modern stellar astrophysics, the knowledge of [$M/H$] is necessary.

A special class of DEBs, which allows for an easier determination of a number of stellar characteristics, are the totally-eclipsing pairs. Geometry requires that components of a totally-eclipsing DEB should clearly differ in radii, which means either different evolutionary phases (e.g. RG+MS) or significantly unequal masses. Depending on the system, the totality phase can last from less than an hour to several days (not taking into account  some extreme and exotic scenarios). During the totality, light of only one star is observed, not being ``contaminated'' by the companion, thus any observations taken at that time are in fact of a single star. This allows for example to: apply simpler analysis techniques, appropriate for single stars; simplify light curve modelling, as some of the degeneracies (e.g. inclination vs. sum of the radii) are less important; or to directly obtain flux ratio and observed colour indices (when multi-band photometry is available) of the components. 

The concept of dedicated observation in total eclipses is not new, but not very often presented, probably because of the difficulties of planning the observations, arising from demanding time constraints. Some examples include studies of late-type components of exotic eclipsing systems, such as post-common-envelope binaries GH~Vir \citep{fulb93} and NN~Ser \citep{haef_nnser} and a $\zeta$Aur-type star 31~Cyg \citep{Bauer_2014}. Other cases were the larger giants in double-RG eclipsing binaries HD~187669 \citep{helm15} and ASAS~J061016-3321.3 \citep{rata21}. More recently, it was realised that totally-eclipsing DEBs are excellent benchmark objects for the upcoming {\it PLAnetary Transits and Oscillations of stars} \citep[PLATO;][]{plato} mission, and programmes aimed at in-eclipse spectroscopic observations have been initiated (P.~Maxted, priv. comm.).

In this publication we present the results of this kind of a dedicated programme, that served as a proof of concept. Its main goal was to show how much our knowledge about a given binary system can be improved with just a single spectrum taken with proper timing. The paper is structured as follows: in Sect.~2 we present the studied targets; Sect.~3 describes the observations of the main project, as well as supplementary data; the methodology is described in Sect.~4, followed by the results and notes for specific systems; Sect.~5 contains discussion of the results; concluding remarks are given in Sect.~6.

\section{Target selection}

This work is a sub-project of a larger undertaking, named Comprehensive Research with \'Echelles on the Most interesting Eclipsing binaries \citep[\creme;][]{helm22}. The main spectrographs used by this project are: CHIRON at the SMARTS 1.5-m telescope in CTIO (Chile), HIDES at the OAO-188 telescope in Okayama (Japan), CORALIE at the 1.2-m Euler telescope in La Silla (Chile), and FEROS at the MPG-2.2m telescope, also in La Silla. Substantial contributions were made with the HARPS (ESO-3.6m, La Silla) and HRS (SALT, South Africa) instruments, and minor contributions with other facilities, including  HDS and IRCS (Subaru, Maunakea), SOPHIE (OHP, France), FIES (NOT, La Palma) or HARPS-N (TNG, La Palma).

The targets for \creme\ were primarily selected from the All-Sky Automated Survey (ASAS) Catalogue of Variable Stars \citep[ACVS;][]{pojm02}. We selected relatively bright ($V<12$~mag) DEBs, suitable for high-resolution spectroscopic observations, and with colour index $V-K>1.1$~mag, suggesting spectral types F and later, with abundant spectral features. For the purpose of this particular study, we searched for total eclipses by visually inspecting the available ASAS-3 light curves. In some cases the ephemerides had to be refined, since the value of the orbital period ($P$) given directly in the ACVS did not produce a satisfactory looking phase-folded light curve. Once the acceptable values of $P$ and zero-phase ($T_0$) were obtained, we estimated the length of the flat part of a total eclipse (\phitot). We used a conservative approach of under-estimating \phitot\ in order to compensate for uncertainties in $P$ and $T_0$. Simultaneously, with ongoing \creme\ observations, we also ruled out a number of triple systems, where the third light could contaminate the spectrum taken in totality.

\begin{table*}[]
    \centering
    \caption{Basic information about the targets and the log of the UVES observations made during total eclipses. The last column indicates which of the two eclipses is the total one: primary (deeper) or secondary (shallower).}
    \label{tab_obslog}
    \begin{tabular}{llrrrrcrrc}
    \hline \hline
     Object & Other ID & RA (J2000) & DEC (J2000) & $V$  & $d_{\rm GDR3}$  & Date of	& Exp. & SNR & Which\\
            &          & [$^\circ$] & [$^\circ$]  & [mag]& [pc]& observation& [s] &  & eclipse? \\
    \hline
AL Ari  & BD+12 378       &  40.6514300 & +12.7355047 &  9.23 & 138.2(4)   & 2014-09-30    & 120 & 108.2 & sec \\
RZ Eri  & HD 30050        &  70.9409591 & -10.6822268 &  7.88 & 197.5(1.0) & 2014-08-21    &  90 & 139.7 & pri \\
A084011 & TYC 8934-2114-1 & 130.0464471 & -65.5289048 & 10.71 & 454.5(2.7) & 2014-09-27    & 570 & 106.3 & pri \\
SZ Cen  & HD 120359       & 207.6462201 & -58.4991937 &  8.59 & 506.8(4.2) & 2014-04-03    & 120 & 114.3 & pri \\
A141235 & TYC 9005-140-1  & 213.1422668 & -60.9733301 & 11.08 & 3374(442)  & 2014-06-28    & 600 &  96.8 & pri \\
A151848 & HD 135671       & 229.6977770 & -56.2778540 & 10.18 & 212.4(8)   & 2014-04-28    & 300 & 121.6 & sec \\
A161710 & BD-18 4254      & 244.2907798 & -18.4671443 & 10.25 & 409.9(2.6) & 2014-06-20    & 330 & 104.2 & pri \\
A180413 & TYC 6251-651-1  & 271.0557684 & -15.9553945 & 10.70 & 1035(19)   & 2014-04-22    & 540 & 118.7 & pri \\
A183946 & BD+05 3897      & 279.9430491 & +05.8954338 & 10.77 & 355(2)     & 2014-05-09    & 540 & 107.4 & sec \\
A191947 & TYC 5717-1347-1 & 289.9456776 & -11.4724449 & 11.44 & 1507(34)   & 2014-08-13    & 720 &  73.6 & pri \\
BQ Aqr  & TYC 6403-563-1  & 354.0371727 & -16.4689819 & 10.69 & 497(4)     & 2014-09-03    & 600 & 100.8 & pri \\
    \hline
    \end{tabular}
\end{table*}

The ACVS targets were supplemented with a number of objects known from the literature to show total eclipses, for which estimating \phitot\ was possible. We note, however, that this literature search was incomplete. In total we identified $\sim$40 objects, mainly from the Southern hemisphere, 11 of which were observed in a dedicated programme with VLT/UVES, and are reported in this work. In eight cases the primary (deeper) eclipse is the total one, suggesting a cooler, inflated secondary (sub-giant or giant), more evolved than the primary. In the remaining three cases, the smaller and cooler secondary hides completely behind the hotter and larger primary, strongly suggesting that both components are main sequence (MS) stars.

Few other notable examples, with totality spectra obtained from non-dedicated \creme\ observations, were already published, for example: HD~187669 \citep{helm15} or ASAS~J061016-3321.3 \citep{deb11}.

\section{Data}

\subsection{UVES Spectroscopy in total eclipses}

Since the spectroscopic observations in totalities have to be taken in a specific, short-time window, we wanted to employ a queue-based instrument installed at a large-aperture telescope, operated in a service mode. We chose the UVES spectrograph \citep{dekk00}, mounted at the VLT~UT2 telescope, located at the ESO's Cerro Paranal observatory (Chile). 

Our observations were made using only the Red arm, cross-disperser \#3, and slicer \#1. This setting gave us the wavelength range of 4779--6808~\AA, with a gap from 5765--5828~\AA. The slit width was set to 0.7 arcsec, which resulted in spectral resolving power of $\sim$57,000. The observations took place during semester P93, between April and September 2014. Observing windows were calculated on the basis of the ephemeris and light curve morphology known at the time, and incorporated as timing constraints in the observing blocks (OBs) submitted to the observatory. Table~\ref{tab_obslog} summarises the UVES observations, together with positions and distances from {\it Gaia} Data Release 3 \citep[GDR3;][]{gaia,gaia_dr3}. 

Spectra were reduced with the pipeline provided by ESO, and the reduced products were downloaded from the ESO Archive. For the analysis we used the 1D-merged, wavelength-calibrated, and barycentric-corrected spectra.  

\subsection{Supplementary spectroscopy}

\begin{table*}
    \centering
    \caption{Observing log of the supplementary CR\'EME and archival ESO spectroscopy.}
    \label{tab_cremelog}
    \begin{tabular}{lcccl}
\hline\hline
 Object	& Instruments & No. of spectra &  From - To & Notes \\
\hline
AL Ari	& HIDES+HARPS+FEROS+CHIRON   & 19+17+10+6 & Sep. 2010 - Nov. 2018 & \citet{grac21}$^{a}$ \\
A084011 & CHIRON+FEROS  & 11+5  & Feb. 2012 - Feb. 2015 &  \\
A141235 & CHIRON+CORALIE    & 10+3  & Feb. 2014 - Jun. 2015 &  \\
A151848 & FEROS & 6 & May 2013 - Aug. 2013  &   \\
A161710 & CORALIE+CHIRON+FIES+HARPS-N   & 12+5+2+1  & Jun. 2012 - Sep. 2015 &  \\
A180413 & CHIRON+HDS    & 7+2   & Aug. 2011 - Oct. 2014 &   \\
A183946 & CORALIE+HIDES & 9+6   & Jun. 2012 - May 2017  &   \\
A191947 & CHIRON    & 8 & May 2014 - Oct. 2014 & \\
BQ Aqr  & CHIRON+CORALIE & 17+10 & Oct. 2010 - Oct. 2013 & \citet{rata16} \\
\hline
\end{tabular}
\\$^a$ Only the HARPS spectra were used in that paper
\end{table*}

Apart from spectra taken in total eclipses, we also report spectroscopic observations used for radial velocity (RV) determination and binary modelling. The majority of the studied systems were included in the \creme\ sample, but also into other researchers' projects \citep[e.g. ][]{grac21}, whose primary goals were different from this work, but related to derivation of stellar parameters.  

The supplementary spectroscopic observations, used for systems from this study, are shortly described in Tab.~\ref{tab_cremelog}. Please note that the publication by \citet{rata16} on BQ~Aqr was in fact based on the \creme\ data. It includes one CHIRON spectrum taken in totality, but the SNR was too low for a proper spectral analysis. For the remaining two systems -- RZ~Eri and SZ~Cen -- we rely on the literature data.
The former was recently observed by \citet{merle}, whose RV measurements we use in this work. We combine them with earlier data from \citet{popp88} to obtain a very long time baseline. The latter dataset was used in previous analyses of RZ~Eri \citep{burk,vive}. The most up-to-date RV data for SZ~Cen are from \citet{ande75}, and were also used in the study by \citet{gron77}.

\subsection{Supplementary photometry}

\begin{table}
    \centering
    \caption{TESS photometry of studied systems used in this work. For SZ~Cen no GI programme is given, because it was observed in long-cadence (1800 or 600 sec) only.}
    \label{tab_tesslog}
    \begin{tabular}{lrcl}
\hline\hline
Object & TIC ID & Sectors & GI Programmes \\
\hline
AL Ari  &  52124859 & 42--44,70 & G04047, G06112 \\ 
RZ Eri  &  56130043 & 5,32 & G011060, G03252 \\
A084011 & 355152640 & 9,10,62--65 & G011083, G05078$^a$ \\
SZ Cen  & 208057430 & 11,38 &  \\  
A141235$^b$ & 332607100 & 38,65 & G03028, G05078 \\
A151848 &  44550552 & 38,65 & G03028, G05078 \\
A180413$^b$ & 360994953 & 80 & G06112 \\
A183946 & 108554368 & 80 & G06112 \\ 
A191947 &   3473481 & 54,80 & G04047, G06112 \\
BQ Aqr  & 402339590 & 2,69 & G011083$^c$,G05078$^a$ \\
\hline
\end{tabular}
\\$^a$ Simultaneously with: G05003.
\\$^b$ Not enough eclipses recorded, orbital period too long.
\\$^c$ Simultaneously with: G011154 and G011048.
\end{table}

Most of our systems were not studied so far, and no dedicated photometry was collected. For those already known from the literature, the observations either come from small-aperture, all-sky surveys (ASAS, SuperWASP), or were taken decades ago. As a follow-up of the CR\'EME project, we systematically apply for the short-cadence photometry through the Guest Investigator (GI) programmes of the Transiting Exoplanet Survey Satellite \citep[TESS;][]{rick14}. Those observations are summarised in Tab.~\ref{tab_tesslog}.

To date, 10 of our systems have space-borne photometric observations available, although SZ Cen in long-cadence only (we used QLP photometry from sectors 11 and 38, taken with 1800 and 600 seconds cadence, respectively). In two cases -- A141235 and A180413 -- the orbital period is much longer than the time spent by the satellite in a single sector, thus no eclipses were recorded for A141235 so far, and only one primary (no secondary) for A180413. One of our systems -- A161710 -- was not in the field of view of TESS at all, although it might be during the Cycle 7. We applied for GI observations of this target, as well as A180413.

The available light curves (LC) are used for binary modelling. Together with RVs, they allow to obtain absolute values of key stellar parameters, including masses $M$ and radii $R$, thus the gravitational acceleration $\log{g}$, which can also be estimated directly from the spectra. Additionally, for short-period, tidally-locked systems (orbital period $P_{\rm orb}$ is the same as the rotational $P_{\rm rot}$), we can also estimate the projected rotational velocity $v\sin{i}$, also retrievable from the spectra. Values of $\log{g}$ and $v\sin{i}$ from the binary modelling can thus be used as a reference for spectral analysis.
In summary, for seven systems we could base our reference $\log(g)$ and $v\sin{i}$ estimates on the TESS data, and for four we could only use the ASAS and ASAS-SN photometry.


\section{Analysis and results}
\begin{figure*}
\centering
    \includegraphics[width=0.95\textwidth]{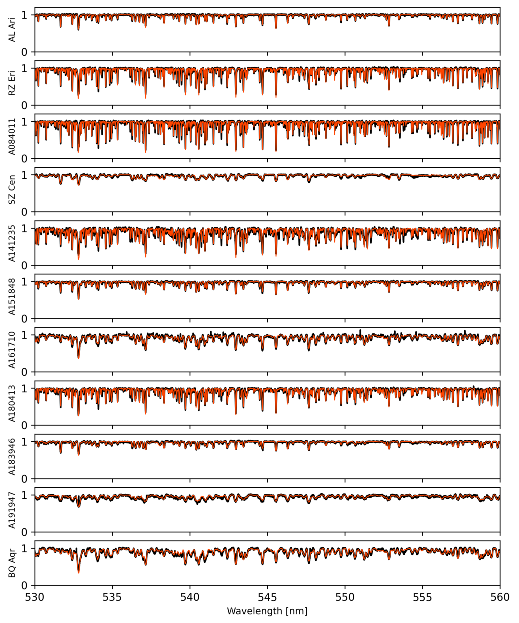}
    \caption{Fragments of spectra of the studied objects: observed, taken during total eclipses (black), and synthetic (red) created with \ispec\ from parameters listed in Tab.~\ref{tab_params_ispec}. All spectra are corrected for the barycentric and radial velocities, and normalised to the continuum. They are plotted in the same scale to emphasise differences between them.}\label{fig_spectra}
\end{figure*}

Below we describe the procedures used in this study, with the focus on analysis of the UVES spectra and application of its results. The binary RV+LC modelling approach, although necessary for this publication, is the same as in previous CR\'EME-based studies, and is here described only briefly. The detailed descriptions of the implemented methodology can be found in previous publications that are based on CR\'EME data \citep[see e.g.][and references therein]{rata16,krishides2017,deb11,helm21,aihya,cht_ayush}. 

\subsection{New and updated eclipsing binary models}

The RVs were measured with our own implementation of the \texttt{TODCOR} routine \citep{todcor}, with synthetic template spectra, and individual measurement errors calculated with a bootstrap procedure \citep{helm12}. The RV solutions were found using the procedure called \texttt{V2FIT} \citep{kona10}, which was used to fit a single- or double-Keplerian orbit to a set of RV measurements of one or two components, utilising the Levenberg-Marquardt minimisation scheme. 

To model the light curves we used the version 40 (v40) of the code \jktebop\ \citep{sout04}, which is based on the \texttt{EBOP} program \citep{popp81}. The out-of-eclipse variations (likely coming from activity) and long-term brightness trends were modelled simultaneously with stellar parameters with the aid of 5-th degree polynomials and sine functions, as described in \citet{helm21}. For long-cadence data of SZ~Cen, we used the option of numerically integrating over a specified exposure time interval (here: 120~s). Parameter errors were estimated with a Monte-Carlo procedure implemented in \jktebop\ (Task 8).

The above scheme was used for all systems except RZ~Eri. We typically use \texttt{V2FIT} and \jktebop\ separately, as the former allows for fitting additional parameters, like differences between zero points of two spectrographs, perturbations of the systemic velocity of a binary (linear, quadratic, periodic), or non-Keplerian effects \citep[see][for more details]{kona10}. For RZ~Eri we decided to fit all the data (LC+RV) simultaneously within \jktebop. The reason was that the solution of \citet{popp88} has an assumed orbital period, given without uncertainties, which does not correspond to the TESS light curve, that is the time between two secondary eclipses (in S05 and S32) was not an integer multiple of the assumed $P$. The TESS curve alone does not have enough information to securely estimate $P$ and orbital eccentricity $e$, as it covers only one primary and two secondary eclipses. By adding the new RV data from \citet{merle} we obtained a time base long enough to overcome those obstacles. A reliable orbital+LC solution could only be obtained by fitting all data sets simultaneously.

The resulting RV and LC solutions are shown in the Appendix, in Figs.~\ref{fig_rvs} and~\ref{fig_lcs}, respectively. The key orbital, stellar, and physical parameters from the RV+LC analysis are shown in Tab.~\ref{tab_paramsRVLC}. Since A141235 and A180413 are single-lined spectroscopic binaries (SB1), absolute values of their stellar parameters were not possible to determine. For previously studied targets, comparison of our results with literature is presented in Tab.~\ref{tab_comparison}.

\subsection{Analysis of spectra in totality}\label{sect_ispec}

\begin{sidewaystable}
\centering
\caption{Atmospheric parameters of the ``spectroscopic'' components of the studied systems, obtained from \ispec\ analysis of the UVES spectra. Values that were held fixed have ``fix'' instead of errors, and those obtained automatically from an empirical relation have ``a''.}\label{tab_params_ispec}
\begin{tabular}{lccccccccccc}
\hline \hline
ID                  & AL Ari    & RZ Eri    & A084011   & SZ Cen    & A141235   & A151848   & A161710   & A180413   & A183946   & A191947   & BQ Aqr \\         
\hline
Component$^a$ & 1 & 2 & 2 & 2 & 2 & 1 & 2 & 2 & 1 & 2 & 2 \\
$T_{\rm eff}$ [K]   & 6540(100) & 4890(170) & 5372(56)  & 7689(94)  & 4520(110) & 6070(66)  & 4844(102) & 5197(78)  & 7630(70)  & 5640(400) & 4630(110) \\      
$\log(g)$           & 4.32(12)  & 2.95(fix) & 3.54(fix) & 3.67(fix) & 1.87(32)  & 4.01(fix) & 2.99(fix) & 2.98(11)  & 4.29(fix) & 3.21(fix) & 3.04(fix) \\      
$[M/H]$             & $-$0.38(5)  & $-$0.29(18) & $-$0.08(5)  & 0.00(5)   & $-$0.25(12) & $-$0.18(6)  & $-$0.42(11) & 0.02(5)   & 0.04(4)   & $-$0.39(29) & $-$0.25(9)  \\      
$[a/Fe]$            & 0.09(4)   & 0.00(19)  & 0.11(5)   & $-$0.13(5)  & 0.18(9)   & 0.03(6)   & 0.16(13)  & 0.04(5)   & $-$0.01(4)  & 0.20(35)  & 0.17(7)   \\      
$v_{\rm mic}$ [km/s]& 1.87(14)  & 1.07(35)  & 1.57(10)  & 5.04(24)  & 1.60(13)  & 1.46(12)  & 2.03(24)  & 1.51(9)   & 3.91(17)  & 2.22(98)  & 1.67(20)  \\      
$v_{\rm mac}^b$ [km/s]& 9.83(a) & 4.09(a)   & 3.59(a)   & ---       & 4.69(a)   & 6.04(a)   & 4.15(a)   & 3.83(a)   &  ---      & 4.17(a)   & 3.93(a)   \\      
$v\sin(i)$ [km/s]   & 17.6(6)   & 10.0(9)   & 7.92(26)  & 55.4(1.2) & 15.4(5)   & 24.6(9)   & 42.6(1.8) & 15.9(3)   & 35.6(7)   & 66.9(8.3) & 48.4(1.8) \\      
\hline 
\end{tabular}\\
\tablefoot{$^a$  Number``1'' means the primary (hotter), ``2'' means the secondary (colder) component. $^b$ Relation not defined for temperatures above 6600 K.}
\end{sidewaystable}

For the sake of clarity, we will henceforth use the following naming convention. The component that is visible during the totality, and whose spectra were recorded with UVES, we call the ``spectroscopic'' one, and the other is referred to as the ``companion''. Their respective parameters will be indexed with ``spec'' and ``comp'', respectively. The spectroscopic component is obviously the larger, and more massive (evolved), but not always the hotter one.

For the determination of atmospheric parameters we used v20230804 of the Python code \ispec\ \citep{ispec14,ispec19}\footnote{\url{https://www.blancocuaresma.com/s/iSpec}}. Before the analysis the UVES spectra were first continuum normalised with \texttt{SUPPNet} \citep{roza22}\footnote{\url{https://github.com/RozanskiT/suppnet}}. Then radial velocities were calculated using the \ispec's module that utilises cross-correlation function, and later used to shift the spectra, so their RVs were equal to 0. We also recalculated the flux errors on the basis of the SNR (see Tab.~\ref{tab_obslog}). In this way we made sure the resulting uncertainties are trustworthy, which was verified by the reduced $\chi^2$, given in the output.

We applied the spectral synthesis method, mostly utilising the code {\tt SPECTRUM} \citep{gra94}, the Model Atmospheres with a Radiative and Convective Scheme (MARCS) grid of model atmospheres \citep{gus08}, and solar abundances from \citet{gre07}. Results were being verified with other choices of those settings, and adopted only if results were consistent. The \ispec\ synthesises spectra only in certain, user-defined ranges, called ``segments''. We followed the default approach, where these segments are defined as regions $\pm$2.5~\AA\,around a certain line. We decided to synthesise spectra around a set of lines carefully selected in such way, that various spectral fitting codes reproduce consistent parameters from a reference solar spectrum \citep{bla16}.

In general, we run fits with the following parameters set free: the effective temperature ($T_{\rm eff}$), logarithm of gravity ($\log(g)$), metallicity ([$M/H$]), $\alpha$-element enhancement ([$\alpha/Fe$]), microturbulence ($v_{\rm mic}$), and projected rotational velocities ($v\sin(i)$). Spectral resolution $R$ was set fixed to the instrument's value of $\sim$57,000, limb darkening coefficient to 0.6, and radial velocity to 0. The macroturbulence velocity $v_{\rm mac}$, which degenerates with rotation, was at all times calculated on-the-fly by \ispec\ from an empirical relation. It is valid only for $T_{\rm eff}<6600$~K, and above that value we set $v_{\rm mac}$ fixed to 0.

From the RV+LC models described in previous section, we obtained the reference values of $\log(g)$ for the spectroscopic components, which we could directly compare to the values found with \ispec. We noted a typical situation where the precision and accuracy of the spectroscopy-based results are worse than for the ``dynamical'' $\log(g)$ from the RV+LC approach. Due the small ``dynamical'' errors, the \ispec-based values of were often outside of $\pm$3$\sigma$ range, and overall agreement was rather poor (occurring mainly thanks to large \ispec\ uncertainties). Therefore, for nearly all SB2 systems -- except AL~Ari -- we decided to repeat the spectral analysis with $\log(g)$ fixed to the RV+LC values. This situation naturally caused concerns about the quality of the results for the SB1 systems. Unfortunately, we are not able to verify them in a way other than checking the consistency of the results through, for example, comparison with isochrones and distance (see the next section).

Results of the \ispec\ fits are presented in Tab.~\ref{tab_params_ispec}. Figure~\ref{fig_spectra} shows the observed UVES spectra with synthetic ones, created in \ispec\ on the basis of the resulting parameters. Table~\ref{tab_comparison}, in the Appendix, shows comparison between various results from this work, including the \ispec\ analysis, and literature. 

Two of the stars we studied had their components analysed spectroscopically before. AL~Ari was presented in \citet{grac21}, where decomposed spectra of both components were analysed with the code {\tt GSSP} \citep{tka_gssp}. The set of fitted parameters was very similar to ours, with the only difference being that the $v_{\rm mac}$ was set free, and the [$\alpha$/Fe] was not taken into account. The overall agreement is very good in all parameters, except for $v_{\rm mac}$ itself, which they found to be much smaller (5.8$\pm$2.0~km/s) than our value taken from an empirical relation (9.83~km/s). The reason most likely is that the effective temperature (6540~K) is close to the range of applicability of the relation. In other aspects, like the results of the RV+LC analysis (Tab.~\ref{tab_comparison}), our results are also in a very good agreement with \citet{grac21}. 

Deconvolved spectra of individual components of BQ~Aqr were analysed by \citet{rata16} with the code {\tt SME} \citep{val_sme}. In that work only the $T_{\rm eff}$ and total metallicity [$M/H$] were fitted, while $\log(g)$ and $v\sin(i)$ were held fixed to the values found in their RV+LC analysis (synchronised rotation was assumed). The $T_{\rm eff}$ of the larger star found by \citet{rata16} -- 4490$\pm$230~K -- is in reasonable agreement with ours (4630$\pm$100~K), although mainly thanks to large uncertainties. The agreement in metallicity is weaker, at the level of 2.6$\sigma$: +0.12$\pm$0.11 from {\tt SME} vs. $-$0.25$\pm$0.09 from \ispec. However, \citet{rata16} calculated with {\tt SME} the overall [$M/H$] without [$\alpha/Fe$], while from \ispec\ we obtained [$M/H$] and [$\alpha/Fe$]. In case of BQ~Aqr we found a substantial enhancements of $\alpha$-elements, which can partially explain the disagreement between our results and Ratajczak's. Most of the other stellar parameters, like the absolute radii or $\log(g)$, are similar for both works (see Tab.~\ref{tab_comparison}), with the exception of $T_{\rm eff,comp}$ ($\equiv T_{\rm eff,1}$).
 
\subsection{Age determination}

\begin{figure}
    \centering
    \includegraphics[width=\columnwidth]{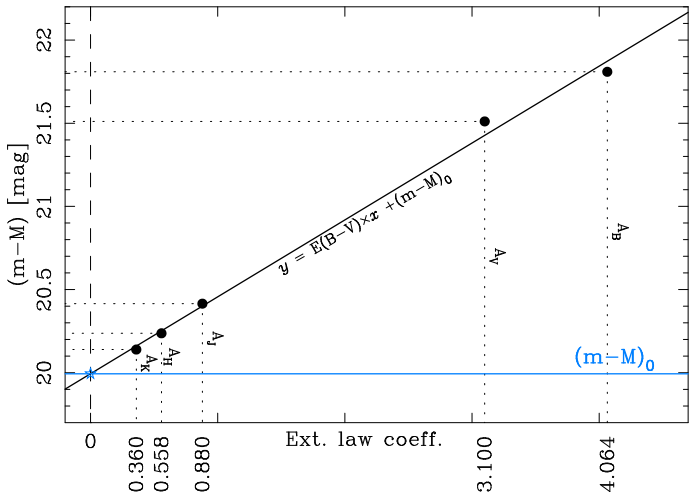}
    \caption{Graphical explanation of the distance and reddening determination method. The distance moduli $(m-M)_\lambda$, calculated from observed magnitudes and synthetic photometry in different bands, are plotted (black dots) as a function of their corresponding law coefficient, by which we define the applied extinction law. The slope and the intercept of the fitted line (solid black) are the values of $E(B-V)$ and reddening-free distance modulus $(m-M)_0$, respectively. The $(m-M)_0$ level is shown in blue. Differences between $(m-M)_0$ and each $(m-M)_\lambda$ represent extinction values $A_\lambda$ in each band (in mag), and are also labelled.}
    \label{fig_dist}
\end{figure}

\begin{figure*}
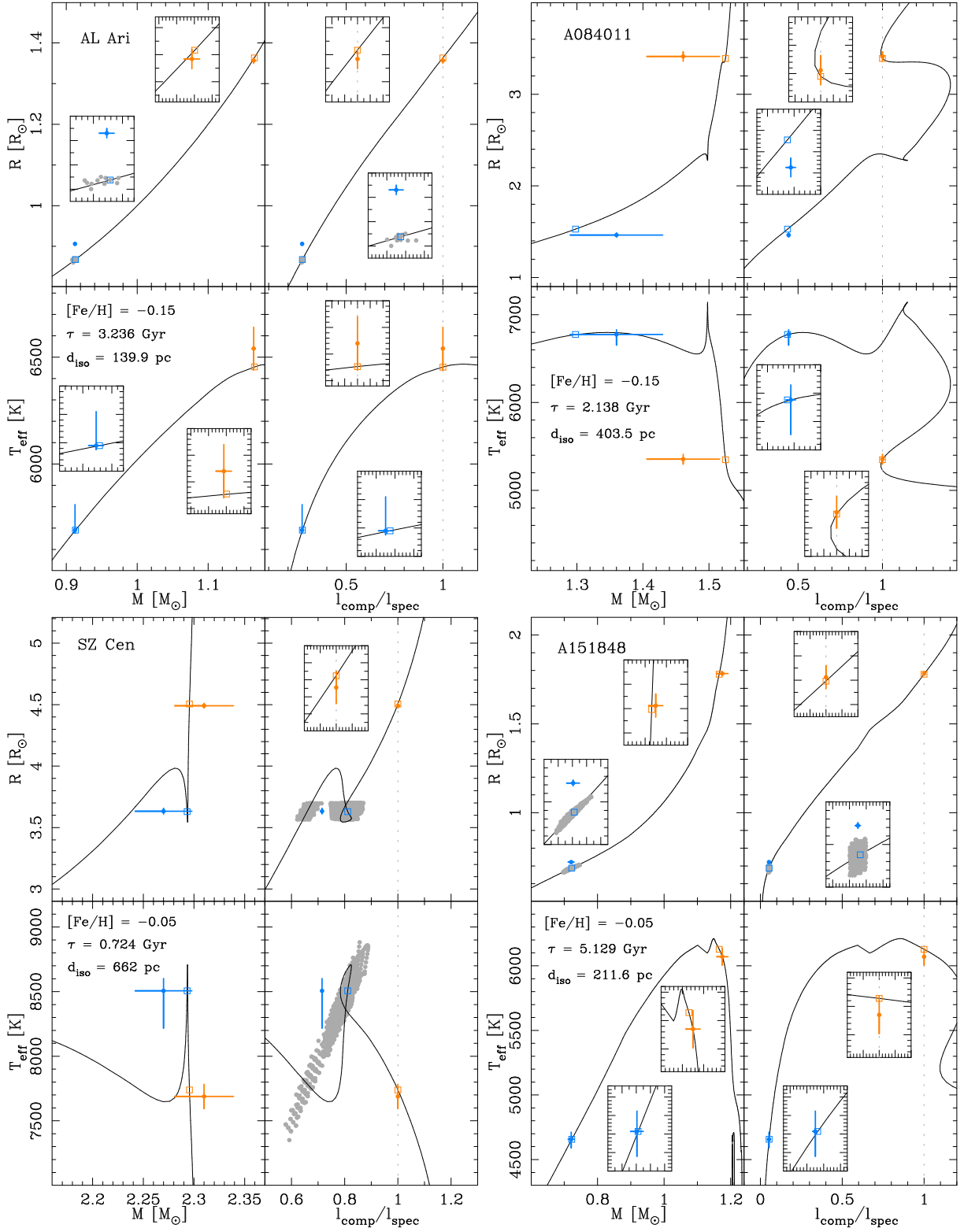

\centering
    \includegraphics[width=0.43\textwidth]{Figures/fig_age_AL_Ari.eps}
    \includegraphics[width=0.43\textwidth]{Figures/fig_age_A084011.eps}
    \includegraphics[width=0.43\textwidth]{Figures/fig_age_SZ_Cen.eps}
    \includegraphics[width=0.43\textwidth]{Figures/fig_age_A151848.eps}
    \caption{Comparison of parameters of SB2 systems with the isochrone (black line) containing the best-fitting model (open square symbols for two components). Age and initial metallicity for the isochrone are given, together with the predicted distance. Measurements and indirect determinations of stellar parameters are marked with filled dots with error bars, and their associated models with open squares. Components named ``spectroscopic'' are represented with orange symbols, while the ``companions'' with light blue ones. For each object we show $M-R$, $l_{\rm comp}/l_{\rm spec}-R$, $M-T_{\rm eff}$ and $l_{\rm comp}/l_{\rm spec}-T_{\rm eff}$ panels. On the $l_{\rm comp}/l_{\rm spec}$ panels the isochrones were scaled with respect to the absolute magnitude of the model that matches the ``spectroscopic'' component, thus it is by definition located on the grey dashed line that marks the unity. In cases where a certain parameter was not used as a constraint in the fitting process (e.g. $R_{\rm comp}$ for AL~Ari or $l_{\rm comp}/l_{\rm spec}$ for SZ~Cen) its model values from all the accepted isochrones are shown with grey dots. Zooms on the vicinity of the determined parameters were drawn for clarity in the insets. In this figure we show results for AL~Ari, A084011, SZ~Cen, and A151848.
    }\label{fig_isos1}
\end{figure*}

\begin{figure*}
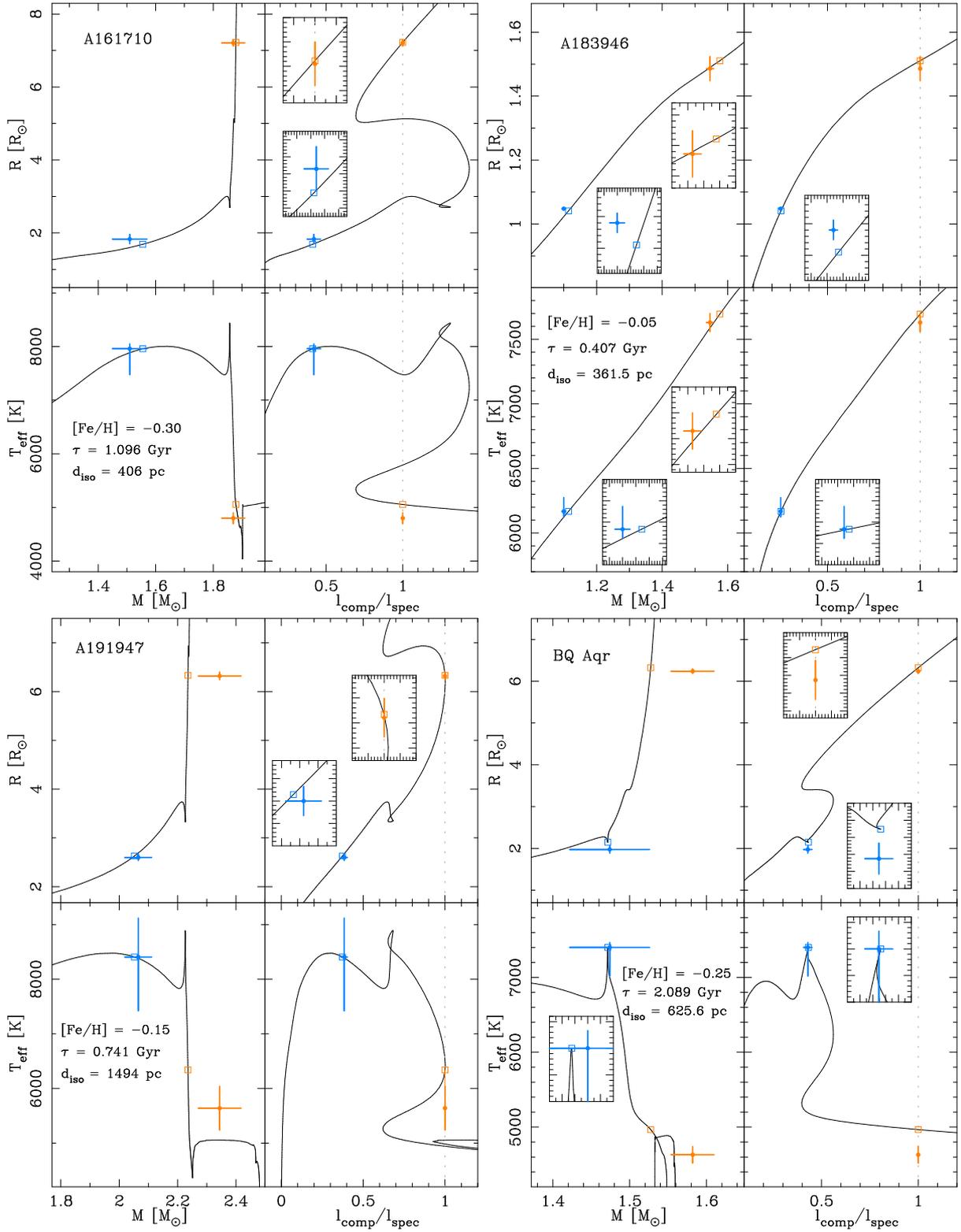

\centering
    \includegraphics[width=0.43\textwidth]{Figures/fig_age_A161710.eps}
    \includegraphics[width=0.43\textwidth]{Figures/fig_age_A183946.eps}
    \includegraphics[width=0.43\textwidth]{Figures/fig_age_A191947.eps}
    \includegraphics[width=0.43\textwidth]{Figures/fig_age_BQ_Aqr.eps}
    \caption{Same as Fig.~\ref{fig_isos1}, but for A161710, A183946, A191947, and BQ~Aqr. Note that for A161710 the $l_{\rm comp}/l_{\rm spec}$ is for ASAS~$V$, not the TESS band. 
    }\label{fig_isos2}
\end{figure*}

\begin{figure}
\centering
    \includegraphics[width=0.48\textwidth]{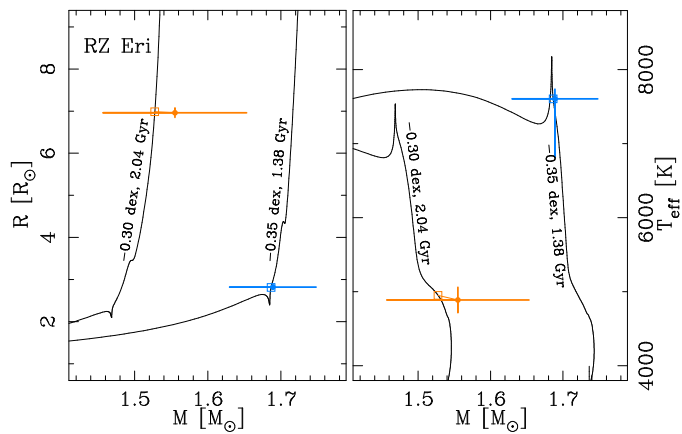}
    \caption{Isochrones fitted to the components of RZ~Eri separately, shown only on the $M-R$ (left) and $M-T_{\rm eff}$ (right) planes. The isochrones are labelled with their age (in Gyr) and initial metallicity. Colour coding for the components is the same as in Figs.~\ref{fig_isos1} and~\ref{fig_isos2}.} \label{fig_isos3}
\end{figure}

\begin{figure}
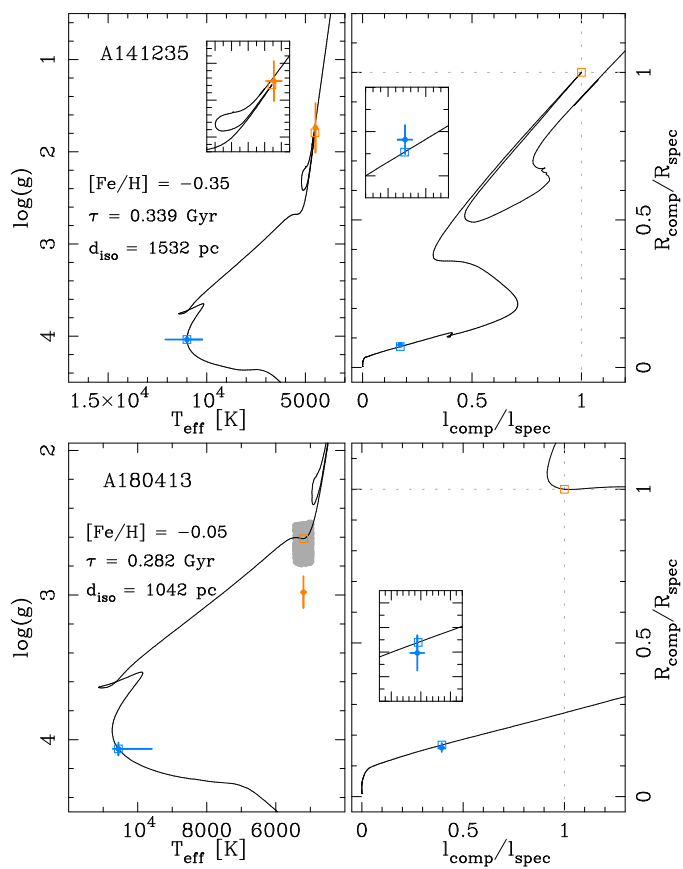

\centering
    \includegraphics[width=0.48\textwidth]{Figures/fig_age_A141235.eps}  
    \includegraphics[width=0.48\textwidth]{Figures/fig_age_A180413.eps}  
    \caption{Same as Figs.~\ref{fig_isos1} and \ref{fig_isos2}, but for SB1 systems (top: A141235, bottom: A180413), and on different planes. We only show $T_{\rm eff}-\log(g)$ (left) and $l_{\rm comp}/l_{\rm spec}-R_{\rm comp}/R_{\rm spec}$ (right). On the latter, the spectroscopic component is shown without errors, and set by definition to (1.0;1.0). Flux ratios refer to the ASAS~$V$ band. 
    }\label{fig_isos_sb1}
\end{figure}

We estimate the ages of our systems using a simple procedure called \texttt{ISOFITTER}, which operates by comparing the results with isochrones, produced using a dedicated web interface,\footnote{\url{http://waps.cfa.harvard.edu/MIST/}} based on the Modules for Experiments in Stellar Astrophysics \citep[{\tt MESA};][]{pax11,pax13,pax15,pax18}, and developed as part of the {\tt MESA} Isochrones and Stellar Tracks project \citep[{\tt MIST} v1.2;][]{choi16,dott16}. We generated a grid that covers ages 10$^{7.0}$ to 10$^{10.2}$~Gyr, in logarithmic scale, every $\log(\tau)$=0.01. For each age we generated isochrones for metallicities between $-$4.0~dex to +0.5~dex, with variable step: 0.05 above $-$0.5~dex and 0.1 below that value. Additionally, we did not include post-AGB phases.

Each isochrone is composed of so-called equivalent evolutionary points (EEPs), taken from evolutionary tracks for a given initial mass. Primary EEPs are a series of points that can be identified in all stellar evolution tracks \citep{dott16}, like the zero-age main sequence (ZAMS), tip of the red giant branch (RGBTip), etc. The secondary EEPs (EEPs henceforth) are chosen to equally split each phase on an evolutionary track into smaller segments. A single EEP can list various information about a model of a star, but the most important for us are: age $\log(\tau)$, initial mass $M_{\rm init}$, current stellar mass $M$, initial metallicity [$Fe/H$]$_{\rm init}$, current metallicity [$Fe/H$], effective temperature $\log(T_{\rm eff})$, gravity $\log(g)$ (which we translate to radius $R$), luminosity $\log(L)$, and synthetic  absolute magnitudes in different bands (we were interested in Johnson-Cousins $U,B,V,R_C,I_C$, 2MASS $J,H,K$, and TESS). Note the distinction between the {\it initial} and {\it current} values of mass and metallicity, which over time may differ significantly. This is due to stellar winds and atomic diffusion, respectively \citep{choi16,dott17}. EEPs with the same age are combined into an isochrone by the MIST web interface. 

We need to introduce the nomenclature we used. With \texttt{ISOFITTER} we searched for the {\it best-fitting model} of a given binary, which means a pair of EEPs of the same age and initial metallicity (i.e. laying on the {\it best-fitting isochrone}) that best reproduce a number of observed and measured stellar properties. We also define {\it accepted isochrones} as those that contain models which meet the assumed criteria -- typically predict values distant at most $3\sigma$ from the measured or observed value. In some cases we also forced other constraints, like the more evolved component must be more massive, or the larger (spectroscopic) component must be strictly cooler or hotter than the companion, depending on the depths of eclipses. Please note that in our approach the best-fitting isochrone contains the best-fitting model, but may also contain some accepted models. 

The goodness of fit of each accepted model was evaluated by a $\chi^2$-like parameter, calculated as
\begin{equation}
    \chi^2 = \sum_{i}{\left( \frac{x_i-x_{0,i}}{\sigma_i} \right)^2},
\end{equation}
where $x_{0,i}$ is the value of parameter $x$ predicted by the given model, $x_i$ is the observed or measured value of this parameter, $\sigma_i$ is its uncertainty, and the summation is done over all the parameters that are constraining the fit. The model with the lowest $\chi^2$ was considered to be the best. It is important to note that we did not make any interpolation between single isochrones of different ages and metallicities, nor between EEPs within one isochrone. Our fits are based purely on a pre-defined grid. 

The uncertainties were also calculated in a simplified way. The most optimal approach in isochrone fitting for single stars would be to use the Bayesian formalism, as for example in \citet{pont04}. In our case it would be quite complicated, because we fit to data for two stars, and for each component we have different measurements (i.e. no spectroscopic data for the companion). We also use constraints based on relative parameters (ratio of fluxes or radii), or observables embedded in a very non-trivial way (observed magnitudes in the distance calculation, see further text). We thus base our uncertainty estimates on the accepted models, that is distribution of the searched parameters among models that meet the fitting criteria. As explained in \citep{pont04}, if the uncertainties of observations/measurements are much smaller than the scale over which the prior of the observed/measured quantity varies, then the probability distribution of the prior can be neglected, and Bayesian approach does not need to be applied. We believe that this is fulfilled in most, if not all, of our cases, thus the non-Bayesian formalism may be applied. Additionally, since our fits are based purely on a pre-defined grid, steps of the grid are included in the uncertainties.

The set of parameters used as constraints in fits varies between SB2 and SB1 systems. In general, for SB2s we used results of the RV+LC analysis: $M$ and $R$ of both components and ratio of fluxes $l_{\rm comp}/l_{\rm spec}$ in the band of the light curve; distance to the system from ${\it Gaia}$, and results of the \ispec\ analysis: $T_{\rm eff}$ and [$Fe/H$] (current) of the spectroscopic component. For the SB1 systems, instead of RV+LC results, we could only use LC-based results, namely the fractional radii of each component $r$~($\equiv R/a$, where $a$ is the major semi-axis of the binary's orbit), and ratio of fluxes. We did, however, add the spectroscopic $\log(g)$ value from \ispec. To calculate $a$, necessary for estimating $r$, we took two stellar masses (from two EEPs), the orbital period $P$ and eccentricity $e$ from the RV or LC analysis, and applied the Kepler's 3rd law. The isochrone-predicted values of flux ratios can be obtained from difference of absolute magnitudes of two EEPs: 
\begin{equation}
l_{\rm comp}/l_{\rm spec} = 10^{-0.4(M_{\rm comp}-M_{\rm spec})}.    
\end{equation}

The isochrone-predicted, reddening-free distance $d_{\rm iso}$ was estimated simultaneously with the reddening $E(B-V)$, using the available observed total magnitudes in different filters, and the predicted total brightness of the system in the same filters, for a given pair of EEPs. Distances in each available band $d_\lambda$ were calculated from $T_{\rm eff}$--surface brightness relations from \citet{ker04}. Then, they were transformed to distance moduli $(m-M)_\lambda$ with the standard relation $(m-M) = 5 \log(d) - 5$. Due to the interstellar extinction and reddening, individual values of $(m-M)_\lambda \equiv (m-M)_0 + A_\lambda$ were obviously not in agreement. To obtain the extinction-free modulus $(m-M)_0$ we fitted a straight line on the $k_\lambda$ vs. $(m-M)_\lambda$ plane, where the $k_\lambda = R_V (A_\lambda/A_V)$ are extinction coefficients in each band. We followed the extinction law of \citet{car89} $A_U:A_B:A_V:A_R:A_I:A_J:A_H:A_K = 4.855:4.064:3.100:2.545:1.801:0.880:0.558:0.360$, which assumes $R_V=3.1$. The slope of the fitted line in this approach is the reddening $E(B-V)$, while the intercept is the extinction-free modulus $(m-M)_0$, which can be translated into the distance $d_0$. 
\begin{equation}
    d_0 = 10^{0.2(m-M)_0 + 1}.   
\end{equation}
A graphical explanation is shown in Fig.~\ref{fig_dist}, on a case where 5 bands were used: $B,V,J,H$, and $K$ (without $U,R$ and~$I$). 

The results of \texttt{ISOFITTER} runs, with all available parameters (with few exceptions that will be described later), are summarised in Tab.~\ref{tab_iso}. For each target we show values of the best-fitting model of the following parameters: age (in Gyr), initial metallicity (in dex), effective temperature of the unseen companion (in K) and isochrone-predicted distance (in pc). For each of those we also list uncertainties, defined as a range of values taken from accepted isochrones, corresponding to the 16th-to-84th percentile of the parameter's distribution. Additionally, for the two SB1 systems we report the masses and radii of their components. For A141235 we obtained:
$M_{\rm spec} = 2.94^{+0.58}_{-0.17}$~M$_\odot$,
$R_{\rm spec} =36.03^{+3.88}_{-3.22}$~R$_\odot$,
for the spectroscopic component, and 
$M_{\rm comp} = 2.58^{+0.36}_{-0.18}$~M$_\odot$,
$R_{\rm comp} = 2.55^{+0.17}_{-0.08}$~R$_\odot$ for the companion. Analogously, for A180413 we obtained:
$M_{\rm spec} = 3.31^{+0.07}_{-0.33}$~M$_\odot$,
$R_{\rm spec} =14.92^{+0.74}_{-2.48}$~R$_\odot$, and
$M_{\rm comp} = 2.65^{+0.10}_{-0.34}$~M$_\odot$,
$R_{\rm comp} = 2.51^{+0.15}_{-0.30}$~R$_\odot$.

We also want to stress that the [$Fe/H$]$_{\rm init}$ of the best-fitting model does not have to coincide with the value found in \ispec, for several reasons. First, we use a grid in [$Fe/H$]$_{\rm init}$ without interpolation. Second, in {\tt MESA/MIST} isochrones and tracks metallicity of a star changes in time, often by more than 0.10~dex. One such example is AL~Ari, for which the resulting [$Fe/H$]$_{\rm init}$ = $-$0.15, while the current value for the spectroscopic component [$Fe/H$] = $-$0.32, as given in the best-fitting model. And finally, the value determined from spectra has its own uncertainty, thus a range of metallicities should be used in isochrone fitting, which is not always the case in literature.

The scheme above has been applied to all our systems, with the exception of RZ~Eri. It has been suggested several times in the literature, that this binary has underwent some major event, namely a significant mass loss from the evolved component, that vastly disturbed its evolution. Therefore, assuming an independent evolution, coevality, and fitting a single isochrone to both components is meaningless. In the case of RZ~Eri, we run \texttt{ISOFITTER} twice, to look for the age of each component separately. Obviously, we only use masses, radii, and metallicity (from \ispec), as well as $T_{\rm eff}$ in one case. No comparison with the GDR3 distance has been made.

In Figs.~\ref{fig_isos1} to \ref{fig_isos_sb1} we show the best-fitting isochrones with our results on various parameter spaces: $M-R$, $M-T_{\rm eff}$, $l_{\rm comp}/l_{\rm spec}-R$ and $l_{\rm comp}/l_{\rm spec}-T_{\rm eff}$ for SB2s (for RZ~Eri only the first two), and $T_{\rm eff}-\log(g)$ and $l_{\rm comp}/l_{\rm spec}-R_{\rm comp}/R_{\rm spec}$ for SB1s. Age and metallicity of the isochrone, and the distance the model predicts, are also given. Errors of our results are shown as $\pm$1$\sigma$ or 16th-to-84th percentile. For some SB2 systems one of the parameters was not used in the fitting process (see the next section), and its values, given by the accepted models, are shown in figures as grey dots.

\begin{table*}
    \centering
    \caption{Results of isochrone fitting. For each system we list the best-fitting isochrone's age and metallicity, effective temperature of the companion, and predicted distance. Each parameter value is accompanied with a range representing 16th-to-84th percentile of distribution of ``accepted'' isochrones (blue points in Fig.~\ref{fig_degeneracy}). RZ~Eri is not shown here, as the results would be meaningless (see text for details). Apart from those parameters, for the two SB1 systems we also evaluated stellar masses and radii, which are given in the text.}\label{tab_iso}
    \begin{tabular}{lcccccccc}
    \hline \hline
Object  & \multicolumn{2}{c}{Age}&  \multicolumn{2}{c}{[Fe/H]$_{\rm init}$} & \multicolumn{2}{c}{$T_{\rm eff, comp}$} & \multicolumn{2}{c}{$d_{\rm iso}$}\\
        & \multicolumn{2}{c}{[Gyr]}&  \multicolumn{2}{c}{[dex]} & \multicolumn{2}{c}{[K]}   & \multicolumn{2}{c}{[pc]} \\
    \hline
AL Ari  & 3.236 & [2.884:3.311] & $-$0.15 & [$-$0.25:$-$0.10] & 5690 & [5676:5810] & 139.9 & [133.3:140.5] \\
A084011 & 2.138 & [2.042:2.291] & $-$0.15 & [$-$0.20:$-$0.05] & 6772 & [6649:6825] & 403.5 & [394.6:414.5] \\
SZ Cen  & 0.724 & [0.700:0.776] & $-$0.05 & [$-$0.15:+0.00] & 8506 & [8217:8602] & 662   & [652:664] \\
A141235 & 0.338 & [0.234:0.417] & $-$0.35 & [$-$0.40:$-$0.25] &11000 &[10220:12100]& 1532  & [1402:1702] \\
A151848 & 5.129 & [4.677:5.623] & $-$0.05 & [$-$0.10:+0.00] & 4660 & [4590:4712] & 211.6 & [209.6:214.3] \\
A161710 & 1.096 & [1.072:1.288] & $-$0.30 & [$-$0.40:$-$0.15] & 7960 & [7476:8052] & 406   & [402:419] \\
A180413 & 0.282 & [0.269:0.363] & $-$0.05 & [$-$0.10:+0.05] &10560 &[9600:10680]& 1042  & [871:1094] \\
A183946 & 0.407 & [0.389:0.468] & $-$0.05 & [$-$0.10:+0.00] & 6170 & [6124:6274] & 361.5 & [359.9:364.1] \\
A191947 & 0.741 & [0.676:0.871] & $-$0.15 & [$-$0.40:+0.20] & 8400 & [7423:9118] & 1485  & [1420:1517] \\
BQ Aqr  & 2.089 & [1.995:2.399] & $-$0.25 & [$-$0.25:$-$0.15] & 7400 & [7020:7467] & 625.6 & [614.4:630.3] \\
    \hline
    \end{tabular}
\end{table*}

\subsection{Notes on individual systems}
\subsubsection{AL Ari}
This is by far the best-studied system in our sample, with very precise physical and stellar parameters available in literature \citep[][and references therein]{grac21}. It was therefore a test-bed for our methodology. We used more RV points (Tab.~\ref{tab_cremelog}), gathered over a much longer time span (8 years), and photometric time series from TESS observations only. On the other hand, the HARPS RVs used by \citet{grac21} are the most precise ones that are available, and their ground-based photometry is multi-band and dates back to 1997, also offering a long time base when combined with RVs. With our data we could not, however, confirm their RV acceleration rate of the baricentre (0.056$\pm$0.011~ms$^{-1}$d$^{-1}$), and found it to be indistinguishable from zero (0.03$\pm$0.04~ms$^{-1}$d$^{-1}$). We do see hints of the apsidal motion (9.4$\pm$7.2~$^\circ$d$^{-1}$), and our slightly longer orbital period confirms the $\dot P=3.8\times10^{-7}$~d\,yr$^{-1}$ value reported in \citet{grac21}.

Both \citet{grac21} and us reached sub-\% precision in stellar masses and radii, together with a very good agreement. Precision in $M$ is about 2 times better in the former work (due to the use of HARPS data only), while we reached lower uncertainties in $R$ (thanks to the TESS photometry). Good agreement is also reached in atmospheric parameters, which is important for understanding potential discrepancies with isochrones. The comparison is shown in Tab.~\ref{tab_comparison} in the Appendix.

The pair is composed of two Main Sequence (MS) stars. The companion (secondary), which is about 10\% smaller and less-massive than the Sun, appears to have its radius overestimated with respect to the models. Such situation is very often observed in active K- and M-type components of DEBs, and is most likely related to rotation and enhanced stellar activity \citep{lop05,char14}. Problems with age determination in such systems are not uncommon \citep[e.g.][]{popp97}. It is also the most likely scenario here, as the TESS light curve of AL~Ari shows small-scale, spot-like modulations outside of the eclipses, and the components appear to rotate synchronously. When looking for a common isochrone, we could not find a satisfactory solution, therefore we decided not to include the secondary's radius as a constraint in the isochrone fitting process (Fig.~\ref{fig_isos1}). Despite that the fit led to a satisfactory solution, and good precision in age determination ($<$10\%).

\subsubsection{RZ Eri} 
This pair is composed of a red giant (RG) spectroscopic component and a companion at the end of its MS evolution or just at the beginning of the sub-giant phase (for simplicity we will call it the MS star). In the RV solution, the latter is formally the more massive star, but, until recently, the large relative errors allowed for solutions where the RG was more massive, as expected in undisturbed evolution scenario. Recent RV data from \citet{merle} showed, however, that the RG truly is the lower-mass star. The RG star is active and spotted, which can be deduced from the out-of-eclipse brightness modulation and asymmetric and variable shape of the secondary eclipse (transit, see Fig.~\ref{fig_lcs}). Such distortions occur when a smaller companion passes in front of cold spots on the surface of the background star \citep{lurie17,helm21}. It clearly affected our results, and hampered the precision in radii down to $\sim$1.6\% level. Still, this is a significant improvement with respect to previous studies \citep[][Tab~\ref{tab_comparison}]{popp88,burk,vive}.

The leading explanation of the lower mass of the RG component is that it has lost a substantial amount of mass some time in the past. \citet{eggy} estimated that it must have been $\sim$20 times the amount expected from stellar winds driven by a magnetic dynamo. This would explain the unexpectedly high extinction towards RZ~Eri, which was shown to be caused by circumstellar, rather than interstellar material \citep{burk}. Another observational features that favour a disturbed evolution are the rotational velocities: the RG seems to rotate slower than the expected pseudo-synchronisation rotation speed\footnote{In case of eccentric orbits the tidal spin-orbit equilibrium occurs for stellar rotation periods different than $P_{\rm orb}$.}, while the MS star spins significantly faster. The latter fact could possibly be explained by a Roche lobe overflow (RLOF) from the spectroscopic component, and thus a transfer of mass. Although for such scenario one needs to take into account the size and shape of the relative orbit ($a=72.6$~R$_\odot$, $e=0.37$).

The observed properties of RZ~Eri, that is the circumstellar material, age inconsistency, and fast rotation of the companion, can also be explained if the companion was a product of a stellar merger. The system would have to have formed as a compact hierarchical triple (CHT), with a short-period inner binary, and a tertiary on a long-period but still relatively tight, significantly eccentric orbit (in line with the currently observed one). Due to dynamical influence of the tertiary (now the RG star), the inner binary's orbit have been tightened and components have coalesced. Mergers in tertiaries have been proposed to explain the observed properties of the binary HD~148937 \citep{frost} or blue stragglers with companions \citep{gleb08}. Numerous CHTs are now being discovered and characterised \citep{mit24,ayush24}, and their dynamical interactions studied \citep{toon20}. Gravitational influence of the outer star can also accelerate the merger \citep{cht_ayush}, especially when it evolves out of the main sequence, and distribution of the angular momenta in the system is changing. Population and statistical studies of mergers in triple systems have been performed \citep[e.g.][and references therein]{kumm23,shar24}, and the current properties of RZ~Eri (e.g. $P$ and $e$) match some of the outcomes, as well as blue straggler binaries found in clusters \citep[see e.g. Fig.~5 in][]{shar24}. The RG star itself could still be largely intact during the process, or has lost some mass through a tertiary or wind Roche lobe overflow (TRLOF or WRLOF). On a $P\sim40$~d ($a~\sim73$~R$_\odot$) orbit it would most likely disturb the material ejected in the collision. 

Study of the merger hypothesis is out of the scope of this paper. We find it, however, an interesting alternative to the enhanced mass loss scenario. Confirmation of the merger would require extensive dynamical and evolutionary simulations, as well as dedicated observations to search for the possible remnants. The tracers of the past merger could be identified in spectra of the MS star or the circumstellar medium, after a careful deconvolution of a series of spectra. Unfortunately, this has not been performed so far, and the data describe by \citet{merle} may be insufficient. The merger remnant could also be visible in radio or sub-mm observations.

In case of RZ~Eri we were looking for two isochrones independently, assuming that evolution of only one component was undisturbed (which may still not be the case). In the RG mass loss scenario, we found that the age of the MS star is $\tau=1.38^{+0.32}_{-0.15}$~Gyr, and [$Fe/H$]$_{\rm init} =-0.35\pm0.15$~dex. Additionally, the evolutionary stage of the companion allows to put some constraints in its effective temperature. Under the RG mass loss scenario we found it to be $7600^{+130}_{-790}$~K. In the merger scenario, the age of the RG is $\tau=2.04^{+0.59}_{-0.46}$~Gyr, and [$Fe/H$]$_{\rm init} =-0.30\pm0.20$~dex. Those two isochrones are presented in Fig.~\ref{fig_isos3}.

Even though in this case we could not show the usefulness of totality spectra for determination of the binary's age, our UVES spectrum and LC+RV solution give important results for this interesting binary. The spectral analysis, which here provided [$Fe/H$] and $T_{\rm eff}$ of the RG star, has been done for this system for the first time, while the obtained masses and radii are the most precise so far. We plan a further investigation of this object.

\subsubsection{A084011} 
This system is composed of a sub-giant (sG) spectroscopic component with a MS companion. Similarly to RZ~Eri, uncertainty in masses is relatively large (5.2 and 3.8\%) but for A084011 it is probably because of systematic errors occurring from a low number of observations around the second quadrature (Fig.~\ref{fig_rvs}). With a couple of additional data points between phases 0.65 and 0.95 it should be possible to improve the mass errors down to sub-\% level. It may be important in the context of the PLATO mission, as A084011 resides in LOPS2, the first field to be observed.

During the isochrone fitting we encountered a systematic discrepancy between isochrone-predicted values of distance ($\sim$403~pc) and the {\it Gaia}~DR3 value (454~pc). With the parameter $RUWE=1.055$, and no entry in the "Non-Single stars" part of the DR3, the {\it Gaia} solution seems to be reliable. But in our fits, no model within the accepted boundaries around the measured stellar parameters, even without spectroscopic ones, has been found to predict distance larger than 445~pc. Therefore, in the final results we decided not to use the distance as a constraint. This discrepancy, although not very strong, remains real and difficult to explain. One possibility is a systematic error in one or two of the observed magnitude values given in {\it Simbad}, which affected the linear fit to the distance moduli. Another explanation might be an overestimated value of [$M/H$] obtained in \ispec. Lower metallicity would mean higher effective temperatures and luminosities, thus the stars should be further away from Earth in order to have the same observed brightness.

\subsubsection{SZ Cen } 
This system hosts the hottest spectroscopic component in our sample, and also one of the most massive. SZ~Cen is also considered a well-studied system, although the RV and LC solutions were obtained in the 70s \citep{ande75,gron77}. The historic RV measurements (Fig.~\ref{fig_rvs}) are all clustered around quadratures, which for a circular orbit is usually sufficient to provide precise and accurate results (here: $\sim$1.2\% for both components). The TESS photometry was available only from the full-frame images (FFI) from sectors 11 and 38, and was extracted with the TESS' Quick Look Pipeline \citep[QLP;][]{qlp}. As can be seen in Fig.~\ref{fig_lcs}, this caused some problems during the modelling. SZ~Cen lays in a field near the Galactic plane ($b\simeq3.5^\circ$), and the QLP severely over-estimated the contamination from background sources (a.k.a. third light). One can note that the eclipses from sector 11 are too deep, both reaching 1~mag, which means that much more than 50\% of the total flux is being blocked twice. Such situation is physically impossible. During LC modelling a {\it negative} value of the $l_3$ parameter had to be used, in order to compensate for the third light over-correction. We found (negative) $l_3$ to be at the level of 42.5 and 9.6\% for sectors 11 and 38, respectively. Fortunately, the crucial parameters ($r_1, r_2, i$, etc.) were in a good agreement between sectors, and the final values of stellar parameters are also in a good agreement with the literature. 

However, one important parameter that was most likely affected was the flux ratio in TESS band. We could not find a satisfactory isochrone fit when $l_{\rm comp}/l_{\rm spec}$ was included. Similarly to A084011, we could also not reproduce the {\it Gaia} distance ($d_{\rm iso}\simeq660$~pc, vs. $d_{GDR3}\simeq500$~pc). The final age and [$Fe/H$]$_{\rm init}$ estimates were thus found without constraints on flux ratio and $d$ (Fig.~\ref{fig_isos1}).

SZ~Cen seems to be composed of an sG spectroscopic component with a main sequence companion, which is just at the end of its MS evolution. Some solutions on the earlier stage, for a bit lower EEP mass, are formally acceptable, but they correspond to significantly lower $T_{\rm eff}$ of the companion, and even negative values of $E(B-V)$. Both our temperature estimates are in general agreement with values obtained by \citet{gron77} from colour indices, which supports our solution. Notably, both \citet{ande75} and \citet{gron77} reported problems with fitting a single isochrone. Our fit could perhaps been better with more precise masses coming from new RV measurements, and a 120~s cadence TESS photometry from postage stamps (SAP or PDC\_SAP), instead of FFIs. Despite the drawbacks, the precision in mass is good, the precision in radii and temperatures is improved with respect to previous studies (Tab.~\ref{tab_comparison}), and the age determination seems to be secure and reliable. The case of SZ~Cen shows that the results from several decades ago may not be enough for the needs of modern stellar astrophysics, and some targets may need to be re-examined.

\subsubsection{A141235} 
This SB1 system has the longest orbital period in our sample, and is possibly the most distant target ($d_{GDR3}$=3374~pc), but the $RUWE$ parameter of its astrometric solutions is quite high ($\sim$2.5). Our accepted models did not reproduce such a large distance ($d_{\rm iso}$=1530~pc), therefore the distance was not included into the set of constraints in the isochrone fit. 

Unfortunately, other parameters, both the spectroscopic and LC-based have relatively large uncertainties, thus they provide rather weak bounds for the models. Still we were able to deduce that the spectroscopic component is an evolved RG, the largest single star in our sample ($R_{\rm spec}=36^{+4}_{-3}$~R$_\odot$), most likely at the tip of the RG branch (TRGB; Fig.~\ref{fig_isos_sb1}). Models with this component being already at the asymptotic giant branch (AGB) are formally not discarded, however less likely, due to the constraint of flux and radii ratios. 

\subsubsection{A151848} 

A151848 is a pair of two MS stars, of G and K spectral types. The (spectroscopic) primary is about to turn into a sub-giant. Both components appear to be active, and the (secondary) companion has the lowest stellar mass in our sample. This feature makes the system potentially interesting for further studies on low-mass stars. As in the case of AL~Ari, the secondary's radius seems to be larger than the isochrones predict ($\sim$6$\sigma$ discrepancy), and thus it was not used as the constraint. Another similarity is the mass of the spectroscopic (more massive) component: 1.175 (A151848) vs. 1.164~M$_\odot$ (AL~Ari). The latter has significantly smaller radius, which suggests a younger age.
Despite the activity, the precision in radii is quite good ($\sim$0.7\% for both components), but in masses is significantly lower (2.0+3.3\%). This is mainly due to low number of RV points (6 per component) and will surely improve with more data. 

Apart from the one radius, the best-fitting model reproduces other parameters quite well. We found a good solution for an age of $\sim$5~Gyr, which makes A151848 the oldest system in our sample (as suggested by the comparison with AL~Ari). 

\subsubsection{A161710} 
This is an SB2 with ASAS photometry only. The (spectroscopic) secondary is a RG at the end of the Hertzsprung gap, while (companion) primary is at the MS stage. Large rotational velocity increases the $rms$ of the RVs and, subsequently hampers precision of the RV-based parameters. The quality of ASAS photometry is far from the optimal, thus the LC-based parameters also have large errors. At the end we reach 3.9 and 2.0\% fractional error in masses, and 7.1+1.2\% in radii, with the lower values for the RG star. Despite the uncertainties, we obtained a good isochrone fit, and secure estimates of age and evolutionary status of both components.

\subsubsection{A180413} 

The TESS photometry available to date for of this binary covered only a single primary eclipse. Therefore, similarly to A141235, for this SB1 system we only used the ASAS photometry, but the RVs were fitted simultaneously with \jktebop, because they much better constrain the small but statistically significant eccentricity. 

The results of the spectroscopic analysis of the totality spectrum allowed for stronger constraints in the isochrone fit than for A141235. The spectroscopic component is clearly a RG star, while the unseen companion is most likely still a MS object. Both companions of A180413 and A141235 actually show very similar values of stellar parameters. Both systems also seem to be in a similar age. The GDR3 distance of A180413 is well established, and our model reproduces it very well. We must note that the $\log(g)$ value of the best-fitting model (2.61~dex) is 3.4$\sigma$ away from the \ispec\ value (2.98$\pm$0.11~dex). It is possible that the \ispec\ value is thus over-estimated. For this reason it has not been used in the final isochrone fit. 

\subsubsection{A183946} 

This is a pair of two MS stars. The spectroscopic component is almost as hot as the one in SZ~Cen. The masses of both stars are well established (0.7+0.5\% fractional error), so are the radii (0.6+0.3\%). The parameters of the (spectroscopic) primary suggested that it lays in $\gamma$~Doradus and near $\delta$~Scuti instability strips. Indeed the TESS light curve revealed pulsations dominated by a single frequency of 9.4255~d$^{-1}$ and amplitude of 0.663~mmag, which fits to the latter category of pulsating stars. Other frequencies were also noted, although with a significantly lower amplitude. The best-fitting model reproduces the measurements and observations in a satisfactory way with quite small age uncertainty.

A183946 is therefore a rare case of a DEB with a $\delta$~Scuti type pulsator, and very precise and accurate determination of many key parameters, including effective temperature, metallicity, and age.

\subsubsection{A191947} 
Another system with a colder sG or RG spectroscopic component, and an earlier type companion. Large rotational velocities and low number of spectra, especially for the companion, were the most likely causes of relatively large errors in masses (3.1+2.2\%). The light curve does not show significant activity-induced variations, so the precision in radii is better (0.9+1.0\%). Of all the parameters obtained with \ispec, the ones of A191947 had the largest uncertainties, which later manifested in large uncertainties in age and [$Fe/H$]$_{\rm init}$ determination. The age-metallicity degeneration is the strongest in our sample, even when the spectroscopic constraints are taken into account (Fig.~\ref{fig_degeneracy}).

The well-established radius of the spectroscopic component is the main constraining factor for its evolutionary phase (it is unlikely that this component is at the horizontal branch) and thus the age of the system. Another strong bound comes from the flux ratio. Its best-fitting model agrees with mass and $T_{\rm eff}$ measurement only at 1.5$\sigma$ level (Fig.~\ref{fig_isos2}). Despite the aforementioned drawbacks, the isochrone-based distance to A191947 is in a good agreement with the GDR3 result.

\subsubsection{BQ Aqr } 

We initially relied on the solution from \citet{rata16}, especially for RV-based parameters, but the recent TESS data from Sector 69, when compared with S02, showed that the orbital period value must be shorter. We found new value of $P$ from TESS S02+S69 data, and repeated the \texttt{V2FIT} analysis on the literature RV measurements.

The spectroscopic component of BQ Aqr is an active sG or RG, while the companion is an MS star, right at the end of this evolutionary phase (as the companion in SZ~Cen). This system shares some general characteristics with RZ~Eri and A084011, but the light curve of BQ Aqr shows the strongest spot-originated modulation in our sample (Fig.~\ref{fig_lcs}). From the shape of the secondary eclipse (i.e. number and location of spot-crossing events) we conclude that there are at least two cold spots, which in their maximum visibility reduce the flux of the active component by $\sim$23\%, and the total flux of the system by $\sim$16\%. With the distorted secondary eclipses and such variation of the flux coming from one of the stars, the LC-based parameters were not found with a good precision. Especially affected were the flux ratio, and the companion's radius. The final fractional errors in radii were 0.9+4.6\% for the spectroscopic and companion components, respectively. Activity, together with high $v\sin(i)$, also hampered the RV measurements of \citet{rata16}, and, subsequently, their and our mass errors (1.8+3.5\% here).

Finding the best-fitting model also turned out to be problematic. None of the accepted models could reproduce the GDR3 distance (497$\pm$4~pc), thus this parameter was not used as a constraint. The adopted model is in 2-3$\sigma$ agreement with several measurements (e.g. $M_{\rm spec}$, $T_{\rm eff,spec}$, $R_{\rm comp}$). 

We note that our adopted model predicts very high temperature of the companion (7400~K), much higher than the result of SME analysis by \citet{rata16} (6390$\pm$230~K; Tab~\ref{tab_comparison}). We suspect the reason for this discrepancy is twofold. First, we obtained significantly lower metallicity, which correlates with temperature in spectral analysis. This has been discussed in Sect.~\ref{sect_ispec}. Second, our determination of the flux ratio may still be inaccurate due to the spots. Without $l_{\rm comp}/l_{\rm spec}$ in the isochrone fit, we get $T_{\rm eff,comp}=6800(220)~K$, which is about 1.3$\sigma$ away from \citet{rata16}. The best-fitting model in this approach predicts $l_{\rm spec}/l_{\rm comp}=3.15(63)$ ($\equiv l_2/l_1$ in Tab.~\ref{tab_paramsRVLC}), [$Fe/H$]$_{\rm init} = -0.15(5)$~dex, $\tau=2.00^{+0.33}_{-0.09}$~Gyr, and $d_{\rm iso}=618(10)$~pc. BQ~Aqr remains a problematic target, without a satisfactory, self-consistent solution.


\section{Discussion}

The main goal of this work is to compare the process of isochrone fitting and age determination in cases when additional spectroscopic information are given, with the situation where they are not (i.e. results from RV+LC only). We did not aim to obtain precise {\it and accurate} stellar ages. This would require the use of several stellar evolution models, and adding some systematic effects into the uncertainty budget. In particular, a number of our systems contain active sub-giant and giant stars, and belong to the RS~CVn variability class. Activity and possible enhanced mass loss \citep[due to stellar winds for instance;][]{eggy} may not be sufficiently implemented into the models we used. Additionally, they may not be well fitted to binary systems, if the binarity affected evolution of one of the components. This might be one of the reasons behind the difficulties in fitting a single isochrone to our measurements or discrepancies with {\it Gaia} distances, which we encountered for several systems (A084011, BQ~Aqr). However, since the same models and measurements were used for cases with and without spectroscopic information, the comparison of those two cases should be reliable, and free of major systematics.

Being aware of the above, we hereby discuss our results.

\subsection{Ages and metallicities with and without spectroscopy}

\begin{figure*}
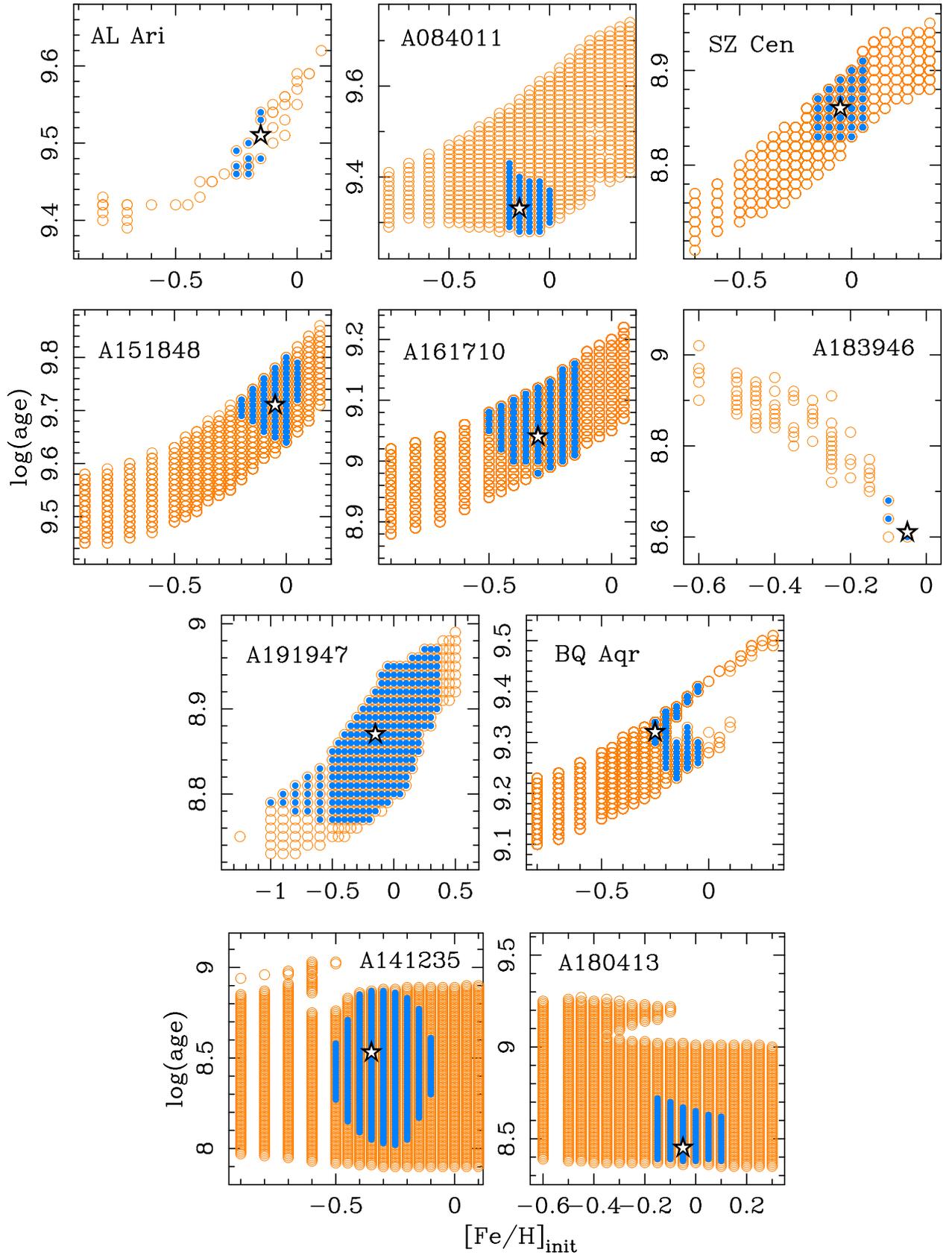

\centering
    \includegraphics[width=0.90\textwidth]{Figures/feh_age_degeneracy_sb2.eps}\vspace{0.5cm}
    \includegraphics[width=0.60\textwidth]{Figures/feh_age_degeneracy_sb1.eps}
    \caption{Graphical presentation of the age-metallicity degeneration in isochrone fitting of the studied systems, on the [Fe/H]$_{\rm init}$ vs. $\log(\tau)$ planes. Orange symbols represent the accepted isochrones in fits constrained only by the RV+LC (for SB2s) or LC-based (for SB1s) parameters. Blue points are accepted isochrones after including spectroscopy into the set of constraints. The white stars mark the isochrones that contain the best-fitting models. Significant reduction in allowed ages and initial metallicities is clearly seen in almost all cases, even when spectroscopic parameters were not well constrained. The SB2 systems are shown on the top three rows (without RZ Eri) , while the two SB1 systems in the lowest one.
    }\label{fig_degeneracy}
\end{figure*}

\begin{figure*}
    \centering
    \includegraphics[width=0.9\textwidth]{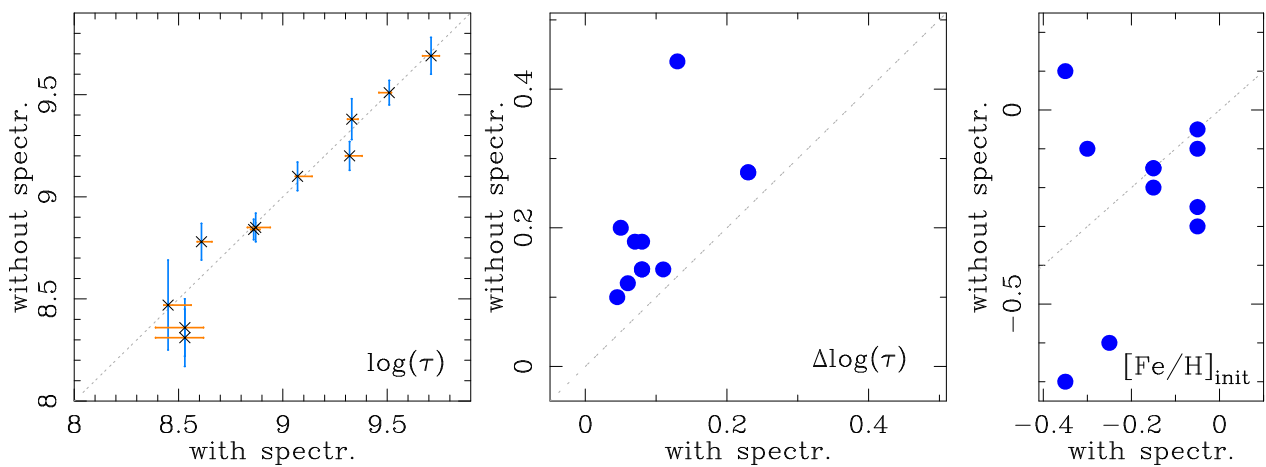}
    \caption{Determination of stellar age through isochrone fitting for cases with and without spectroscopic constraints. The grey dotted line on each panel shows the equality relation ($y=x$). {\it Left:} Logarithm of age ($\log(\tau)$). Uncertainties from each approach are shown with different colours for easier comparison. {\it Centre:} Uncertainties in age $\Delta\log(\tau)$ defined as 16-to-84th percentile or $\pm$1$\sigma$ ranges. {\it Right:} Comparison of the best-fitting initial metallicities [$Fe/H$]$_{\rm init}$. }
    \label{fig_nosp_isp}
\end{figure*}

To directly show the usefulness of having independent estimates of metallicity and effective temperatures, for example from spectra taken in totality, for each system (with the exception of RZ~Eri) we run isochrone fits in two cases: without and with results from \ispec\ analysis. In both cases distance was also taken into account. Results are shown in Fig.~\ref{fig_degeneracy}, on a [$Fe/H$]$_{\rm init}$ vs. $\log(\tau)$ plane. 

Isochrones, for which models were found to be in a 3$\sigma$ agreement with measured stellar masses, radii (if SB2), fractional radii (if SB1), and flux ratio only, are shown in orange. Blue points represent isochrones that contain models with additional 3$\sigma$ agreement with metallicity and effective temperature of the larger component, plus its $\log(g)$ for SB1s, as determined from UVES spectra. One can see, especially for SB2s, that the orange points form bands, showing strong degeneracy between the age and metallicity. This means that having only masses, radii, and ratio of fluxes is not sufficient for meaningful age determination, even though $l_2/l_1$ is very well determined in systems with total eclipses. Addition of distance often helps in ruling out some extreme values, but does not solve the degeneracy completely. With information about only one component, its $T_{\rm eff}$, [$Fe/H$] (note it may be different for each star), and $\log(g)$ (if unknown before) we can narrow down the range of possible ages, even if the errors in atmospheric parameters are relatively large (e.g. A191947). It's like reducing something of a size of a bottle of wine to the size of its label. Cases like A084011, SZ~Cen, or A161710 show that not only [$Fe/H$] but also temperature helps to solve the degeneracy, as there are isochrones for acceptable metallicities, but which not met other criteria. 

The improvement in age determination is shown more qualitatively in Fig.~\ref{fig_nosp_isp}. We compare ages found with the use of $T_{\rm eff,spec}$ and [$Fe/H$]$_{\rm spec}$ ($X$ axis) with those found based on RV+LC parameters (and distance) only ($Y$ axis). The three ``youngest'' points are our two SB1 systems (A141235 shown twice, see further text). The ages themselves are generally in a good agreement (left panel). The central panel directly compares uncertainties from both approaches, where we plot ranges of the age distribution of accepted isochrones corresponding to their 16-to-84th percentile ($\pm$1$\sigma$ for normal distribution). It shows that if the spectroscopic data are used, age is always better determined. The most striking example is the SB1 system A180413, with 0.44~dex reduced to 0.13~dex in $\log(\tau)$. The best SB2 example is A084011 with 0.20~dex uncertainty reduced to 0.05.

We note that similar behaviour can also be seen for most of the other parameters, such as $T_{\rm eff,spec}$, $T_{\rm eff,comp}$, and also others not taken into account in the fit (i.e. $R_{\rm comp}$ for AL~Ari). The best fits with and without spectroscopy are mostly similar, but with significantly smaller errors in the former. This also applies to $T_{\rm eff,spec}$ determined from \ispec\ and purely from isochrone fit. 

The exception to this rule is initial metallicity, which can be seen in the right panel of Fig.~\ref{fig_nosp_isp}. The results without spectroscopy can be very much different from those obtained with it If they agree, it is most likely accidental. Two stars of the same mass and different metallicities should have different effective temperatures, luminosities, and hence absolute magnitudes. Thus, in principle, the distance to a star (or a binary in our case) could help to indirectly constrain the $T_{\rm eff}$ and [$Fe/H$]. In general, we do see such behaviour in our results, but even with constraints on $d$, the [$Fe/H$]$_{\rm init}$ is not very well determined (typically with $\pm$0.2-0.4~dex uncertainty).

A141235 is an interesting case, as without spectroscopic constraints we found two solutions of equal $\chi^2$: for ([$Fe/H$]$_{\rm init}, \log(\tau)$) = ($-$0.7, 8.36) and (+0.1, 8.31). Between these two one can find many models with only slightly worse quality of the fit. This is of course due to very loose constraints coming from the quality of the photometric data (ASAS) and the fact that the GDR3 distance was not taken into account. Despite that, the age seems to be determined with good precision, however, still under-estimated ($\log(\tau)$=8.53 from the full fit).

\subsection{Spectra in totality vs. disentangled}
The scope of this work was to show the usefulness of spectra taken during total eclipses (through dedicated and carefully programmed observations) in studies of eclipsing binaries, in particular in determining the age by obtaining an independent estimate of temperature and metallicity of one of the components. This is obviously limited only to the totally eclipsing systems. There are, of course, other means to achieve the same goal from a set of individual spectra of any SB2 (not only totally-eclipsing), taken in different orbital phases, when lines of both stars are seen. If the fluxes of both stars are roughly comparable, both components can thus be successfully analysed. For example, the \texttt{GSSP} code \citep{tka_gssp} includes a ``composite'' mode, where atmospheric parameters of both components are found simultaneously from a single observed spectrum. The most popular approach, however, is separation of individual components from a number of composite spectra through some sort of a disentangling technique \citep[a.k.a. deconvolution, decomposition, e.g.:][]{bagn91,simo94,hadr95,ili_fd3,kona10,tka_lsd}.

In several cases where both disentangled and totality spectra were used \citep{helm15,rata21}, the resulting atmospheric parameters are in a very good agreement. The disentangled spectra typically have higher SNR, because are constructed of a number of single observations -- the more composite spectra, the higher the SNR of the deconvolved ones. Another advantage of disentangling is the ease and flexibility of planning the observations. The spectra can be taken at almost any time during the orbital period, as long as they cover different orbital phases and RV values of the components, and there is a sufficiently large number of them. With high SNR of the individual spectra ($>$70) one can sometimes obtain satisfactory deconvolution after as few as 5-6 visits \citep{helm21,mill22}, although some methods require minimum of 8 \citep{kona10}. 

On the contrary, planning the observations for totality is difficult because of how short can it last. For some short-period systems the window of opportunity lasts only single hours\footnote{Some observatories, offering queue or service mode observations, do not accept programs with short windows.}, although it opens up many times per observing season, every orbital period. It may however be challenging to have the telescope ready for the precise moment, or gather sufficient SNR in a relatively short time. For long-period binaries, the totality can last up to few days, but it appears every couple of months or rarer. It is easy to imagine the loss of such an opportunity due to weather or technical problems. It was most likely for these planning and time constraining reasons that our original program on UVES was not carried out completely. 

The huge advantage comes, however, with the telescope time -- a single visit during the eclipse is enough, while disentangling requires multiple observations, still with a sufficient SNR (we want to note that for meaningful RV measurements a SNR$\sim$20-30 is often enough, while spectra that will undergo disentangling should optimally have SNR$>$50). The further data analysis is also simpler, as the disentangling is itself a complicated process, prone to systematic errors (e.g. from incorrect continuum normalisation) that further affect the results. Disentangling is mathematically an under-determined problem, meaning that there are more variables than data, and various results can reproduce the given input. The products are thus affected by systematic trends, which need to be carefully corrected for \citep{ili_fd3}. Then the decomposed spectra need to be re-scaled, basing on the flux contribution from each component, or the flux ratio, which itself is wavelength-dependent and does not have to coincide with the flux ratio obtained from the LC fit. Improper re-scaling will lead to incorrect depths of spectral lines. 

A quick solution to this is actually to take a spectrum during the total eclipse with the same instrument and settings. The line profiles and depths are thus known for one of the components, which makes the whole spectral decomposition process much easier and faster. Totality spectra can thus effectively support and simplify the deconvolution.

In summary, we can conclude that even on its own, without other spectroscopic observations or without a full RV+LC model of the binary, a totality spectrum provides valuable information about the spectroscopic component, and the system as a whole. The optimal observational strategy for comprehensive studies of totally-eclipsing DEBs, that balances between minimal telescope resources and maximal scientific output, could thus be as follows:
\begin{itemize}
    \item Obtain one totality spectrum with high SNR ($\gtrsim$80).
    \item Obtain a few (4-6) out-of-eclipse spectra with the same instrument and somewhat lower SNR ($\sim$40-60) for disentangling.
    \item Obtain many more ($>$10) observations with even lower SNR ($\sim$20-30), possibly with smaller telescopes and lower-resolution spectrographs, for RV measurements. 
\end{itemize}

\section{Conclusions}
In this work we presented 11 totally-eclipsing DEBs, which we modelled with available spectroscopic and photometric data, and which were observed with VLT/UVES during their total eclipses. The UVES observations allowed for direct and independent characterisation of one of the components, which we called ``spectroscopic'', and putting strong constraints on the nature of the second component (``companion'') and the system as a whole. Majority of our targets did not have, till now, their ages nor their stellar and orbital parameters determined. 

We found that with good spectral analysis results for at least one star of a binary, one can significantly lower the uncertainty in age estimation, assess the initial metallicity of the whole system, and even point out stellar parameters with problematic or uncertain results, like {\it Gaia}~DR3 distances, or radii of inflated components. Most of our age determinations have precision better than 10\%, even down to 5\% for the best case. We also conclude that without spectroscopic constraints the metallicity can not be securely estimated with isochrones, even with high-precision distance and quality RV+LC analysis.

\begin{acknowledgements}
We would like to thank the anonymous Referee for invaluable comments and suggestions.

We would like to thank the staff of the observatories involved in this work, especially the VLT/UVES, for conducting the challenging, yet successful observations. KGH would like to thank Dr. Pierre Maxted from the School of Chemical and Physical Sciences of the Keele University, and Dr. Nikki Miller from the Nicolaus Copernicus Astronomical Center of the Polish Academy of Sciences (NCAC PAS), who independently picked up a similar idea, for discussions on the topic. We would also like to thank Dr. Tomasz Kami\'nski, Thomas Steinmetz, and Ayush Moharana (all from NCAC PAS) for fruitful discussions.

KGH is supported by the Polish National Science Center through grant no. 2023/49/B/ST9/01671.

This project is a result of the Summer Student Program organised by the NCAC~PAS.

This research made use of data collected at ESO under programmes 082.D-0933, 083.D-0549, 084.D-0591, 085.D-0395, 085.C-0614, 086.D-0078, 088.D-0080, 090.D-0061, 190.D-0237, 091.D-0145, 091.D-0414, 093.D-0254, 094.A-9029, and 094.D-0056, 
as well as through CNTAC proposals 2011B-021, 2012A-021, 2012B-036, 2013A-093, 2013B-022, 2014A-044, 2014B-067, 2015A-074.
The observations on NOT/FIES and TNG/HARPS-N have been funded by the Optical Infrared Coordination Network (OPTICON), a major international collaboration supported by the Research Infrastructures Programme of the European Commissions Sixth Framework Programme. OPTICON has received research funding from the European Community's Sixth Framework Programme under contract number RII3-CT-001566.
This research is based in part on data collected at the Subaru Telescope, which is operated by the National Astronomical Observatory of Japan. We are honoured and grateful for the opportunity of observing the Universe from Maunakea, which has the cultural, historical, and natural significance in Hawaii.
This paper includes data collected by the TESS mission, which are publicly available from the Mikulski Archive for Space Telescopes (MAST). Funding for the TESS mission is provided by NASA’s Science Mission directorate. This work has made use of data from the European Space Agency (ESA) mission {\it Gaia} (\url{https://www.cosmos.esa.int/gaia}), processed by the {\it Gaia} Data Processing and Analysis Consortium (DPAC, \url{https://www.cosmos.esa.int/web/gaia/dpac/consortium}). Funding for the DPAC
has been provided by national institutions, in particular the institutions participating in the {\it Gaia} Multilateral Agreement. {\color{white} https://www.youtube.com/watch?v=ysUjYAi0WcQ}

\end{acknowledgements}

   \bibliographystyle{aa} 
   \bibliography{totals} 

\begin{appendix}

\section{Models of studied systems.}

\begin{figure*}
    \centering
    \includegraphics[width=0.33\textwidth]{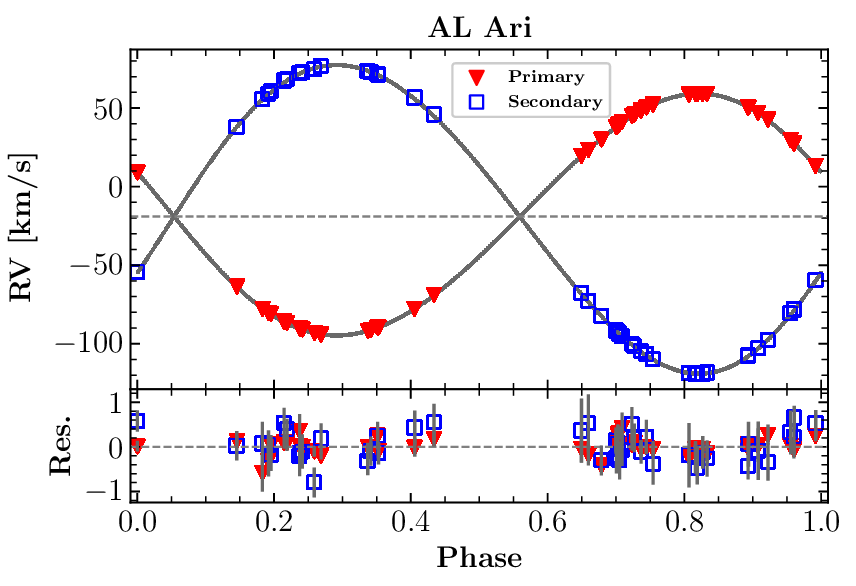}     %
    \includegraphics[width=0.33\textwidth]{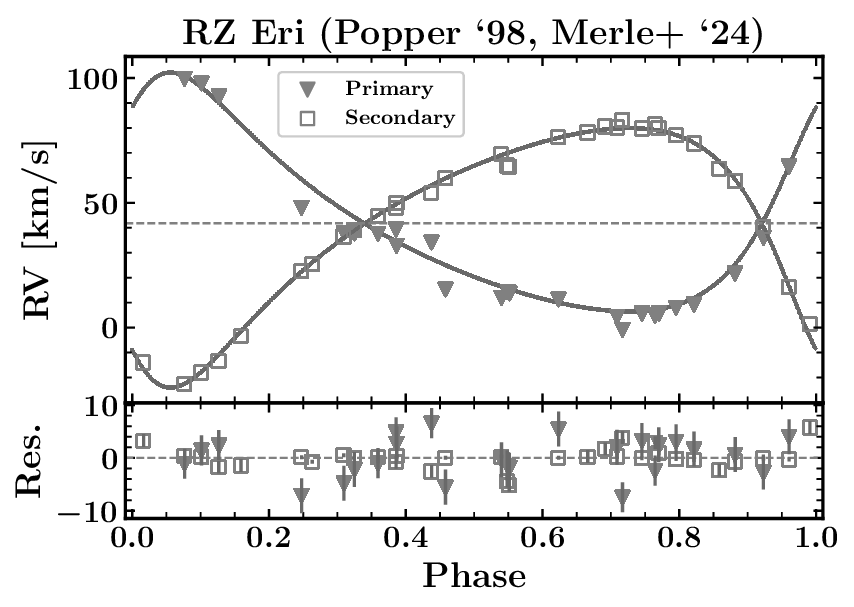}    %
    \includegraphics[width=0.33\textwidth]{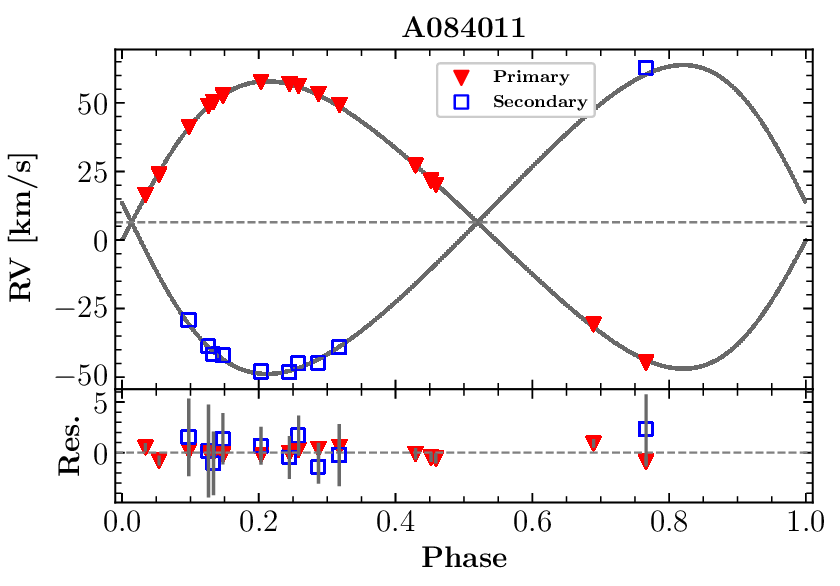}   %
    \includegraphics[width=0.33\textwidth]{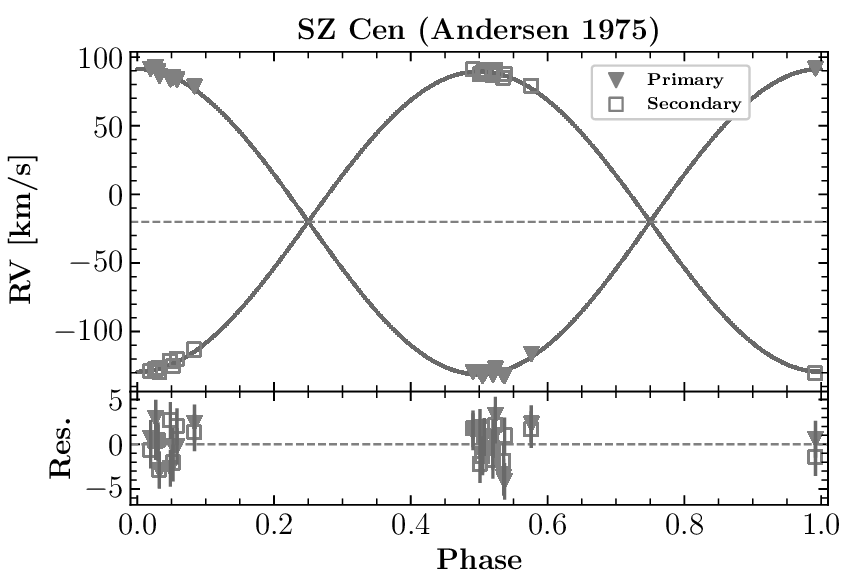}    %
    \includegraphics[width=0.33\textwidth]{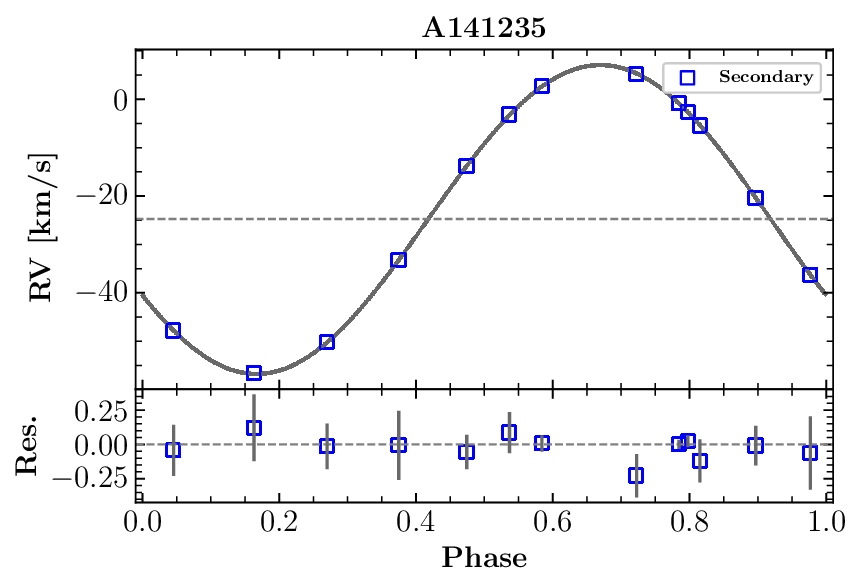}   %
    \includegraphics[width=0.33\textwidth]{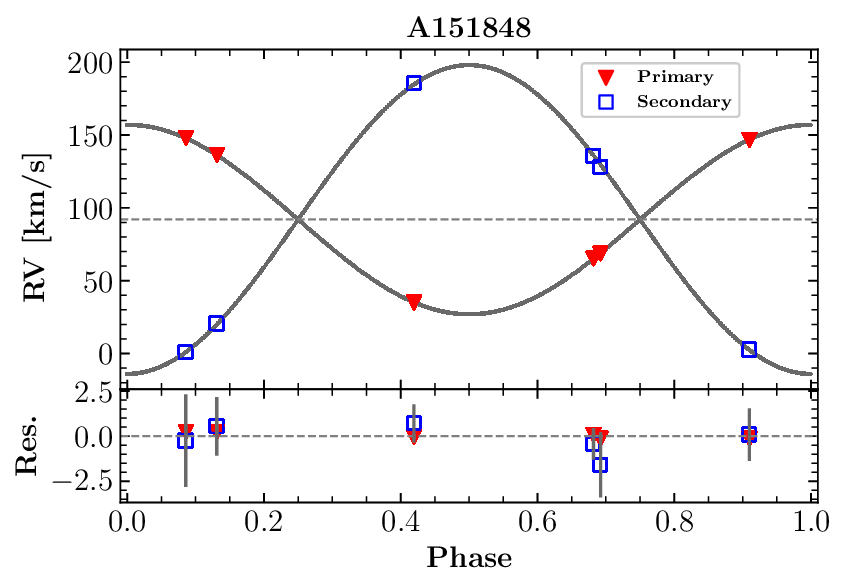}   %
    \includegraphics[width=0.33\textwidth]{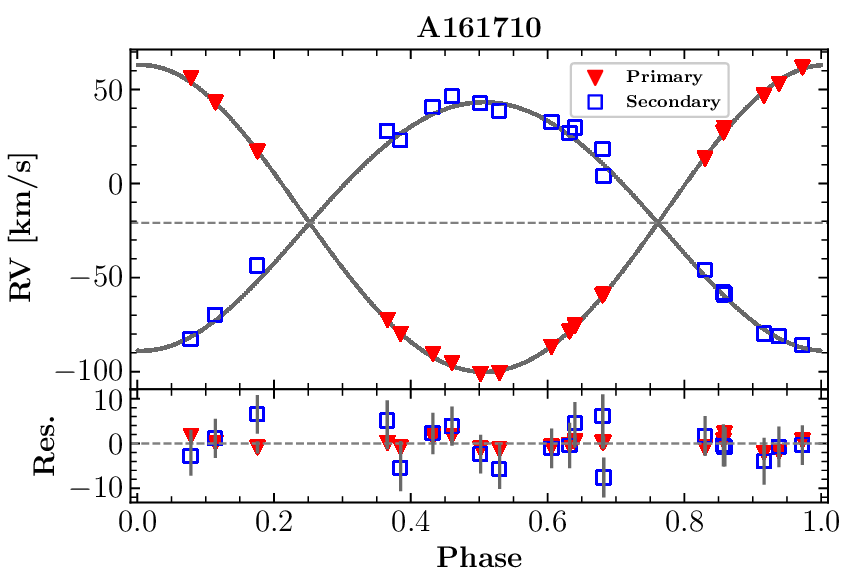}   %
    \includegraphics[width=0.33\textwidth]{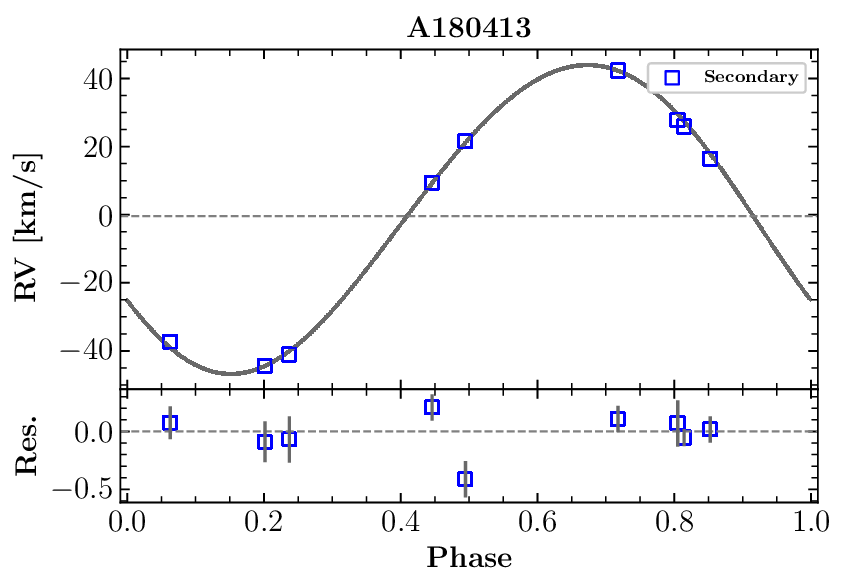}   %
    \includegraphics[width=0.33\textwidth]{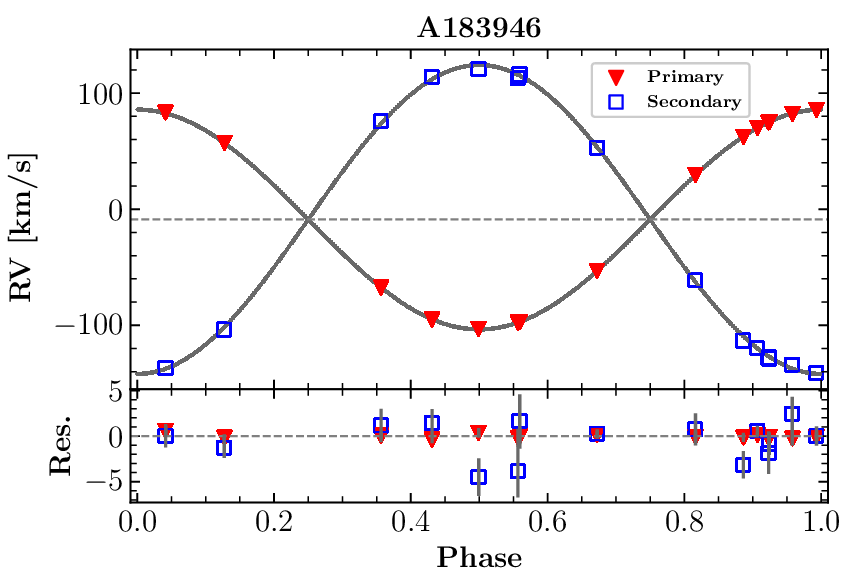}   %
    \includegraphics[width=0.33\textwidth]{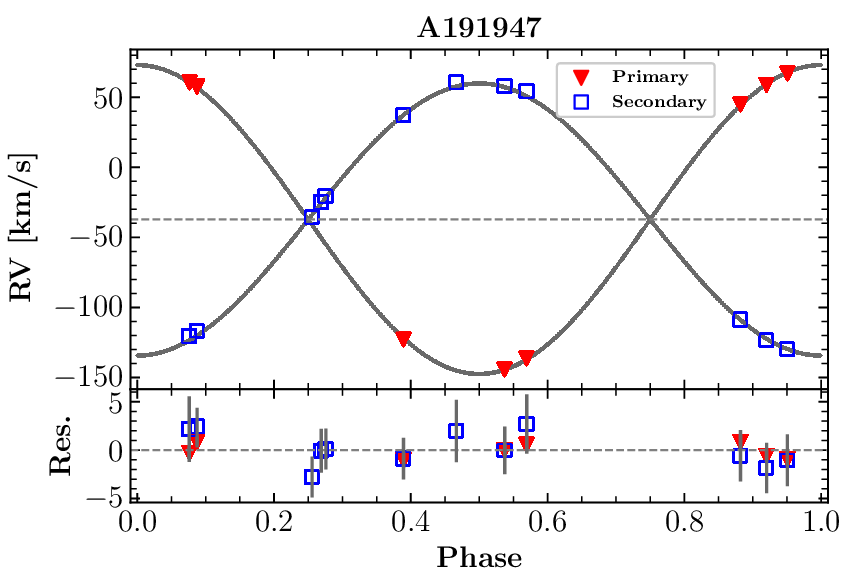}   %
    \includegraphics[width=0.33\textwidth]{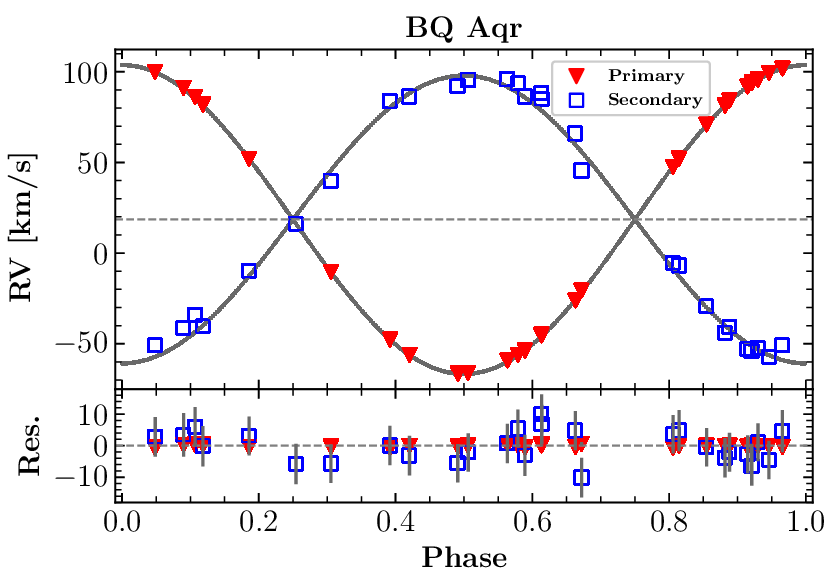}     %
    \caption{Radial velocity solutions based on the CR\'EME (red and blue points) and literature (grey) measurements. Phase 0 is set to $T_P$ -- the time of pericentre passage (for $e>0$) or quadrature (for $e=0$). A141235 and A180413 are SB1 systems with secondary (cooler) components visible. Solution for RZ~Eri was re-done using RVs from \citet{popp88} and \citet{merle}, and TESS photometry. The panel for SZ~Cen  was created from the original values taken from \citet{ande75}. The observations of BQ~Aqr are the same as in \citet{rata16}, but the model has been refined.}
    \label{fig_rvs}
\end{figure*}

\begin{figure*}
    \centering
    \includegraphics[width=0.33\textwidth]{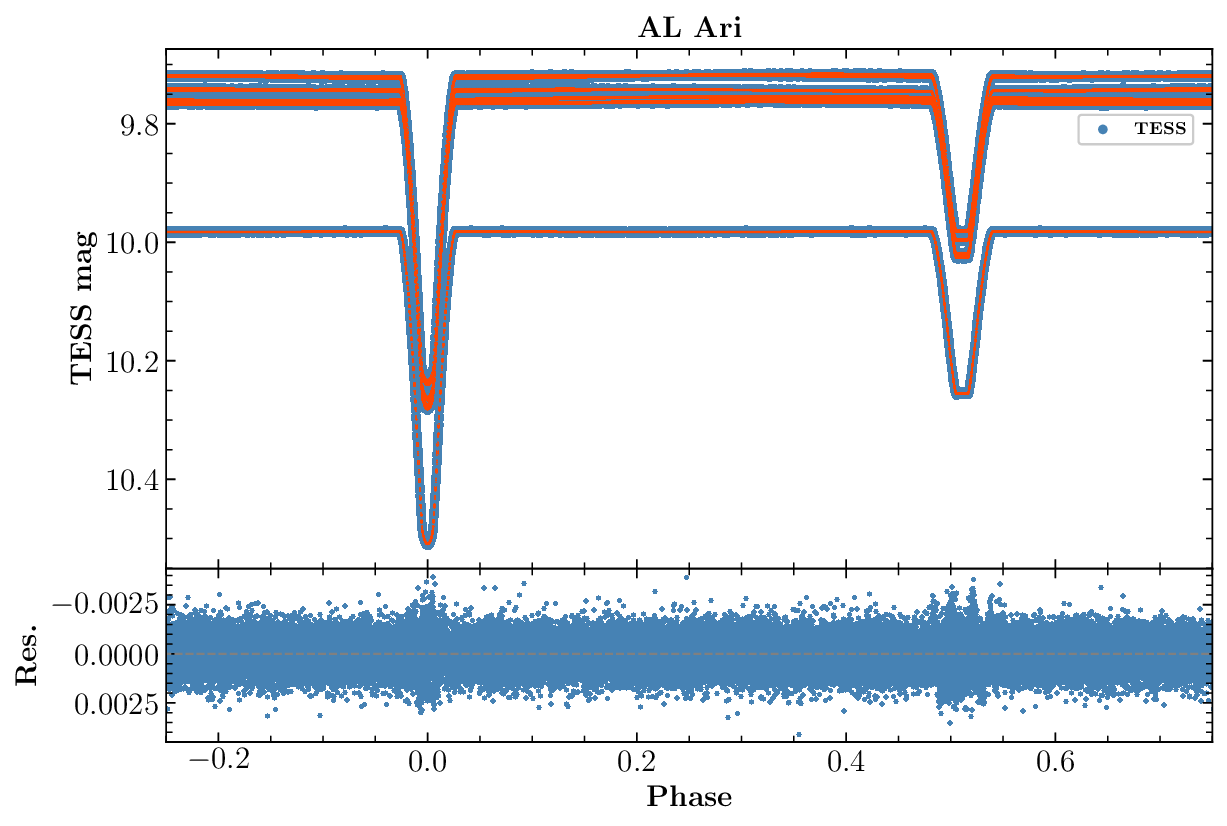}   %
    \includegraphics[width=0.33\textwidth]{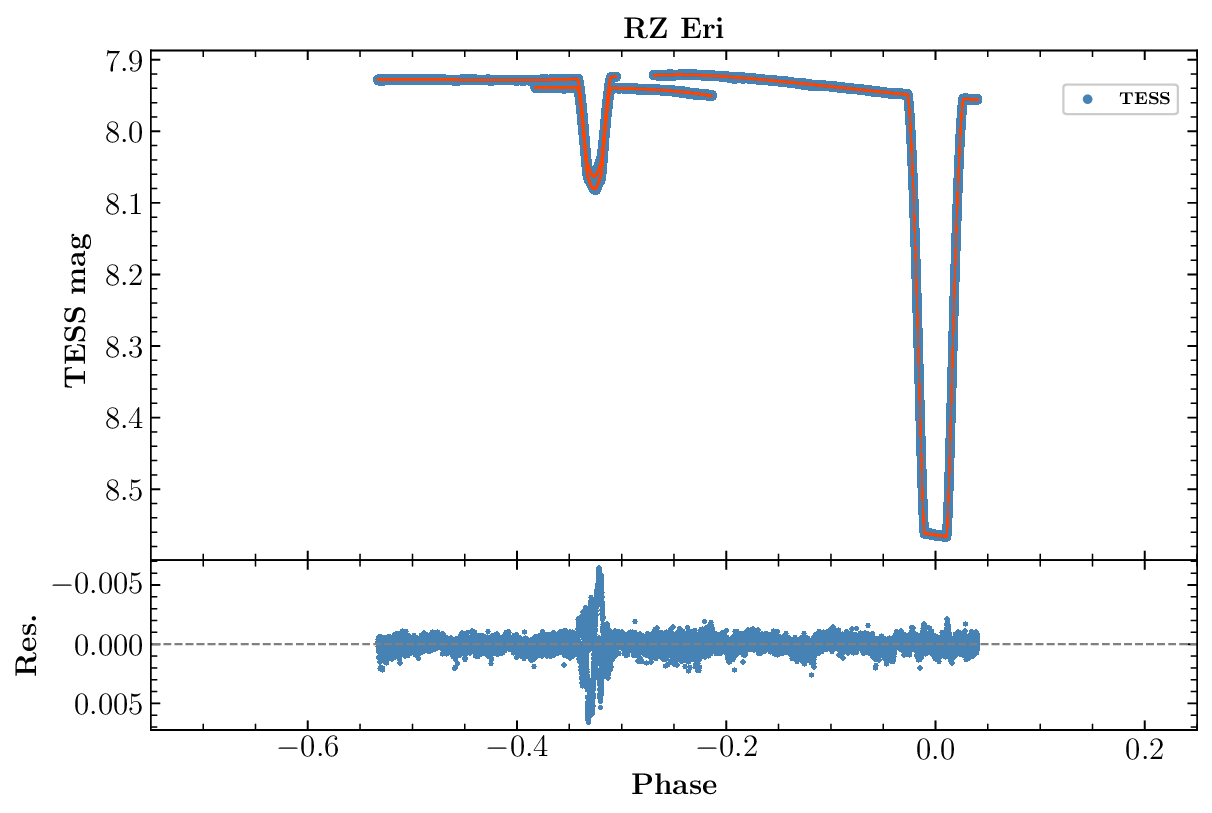}   %
    \includegraphics[width=0.33\textwidth]{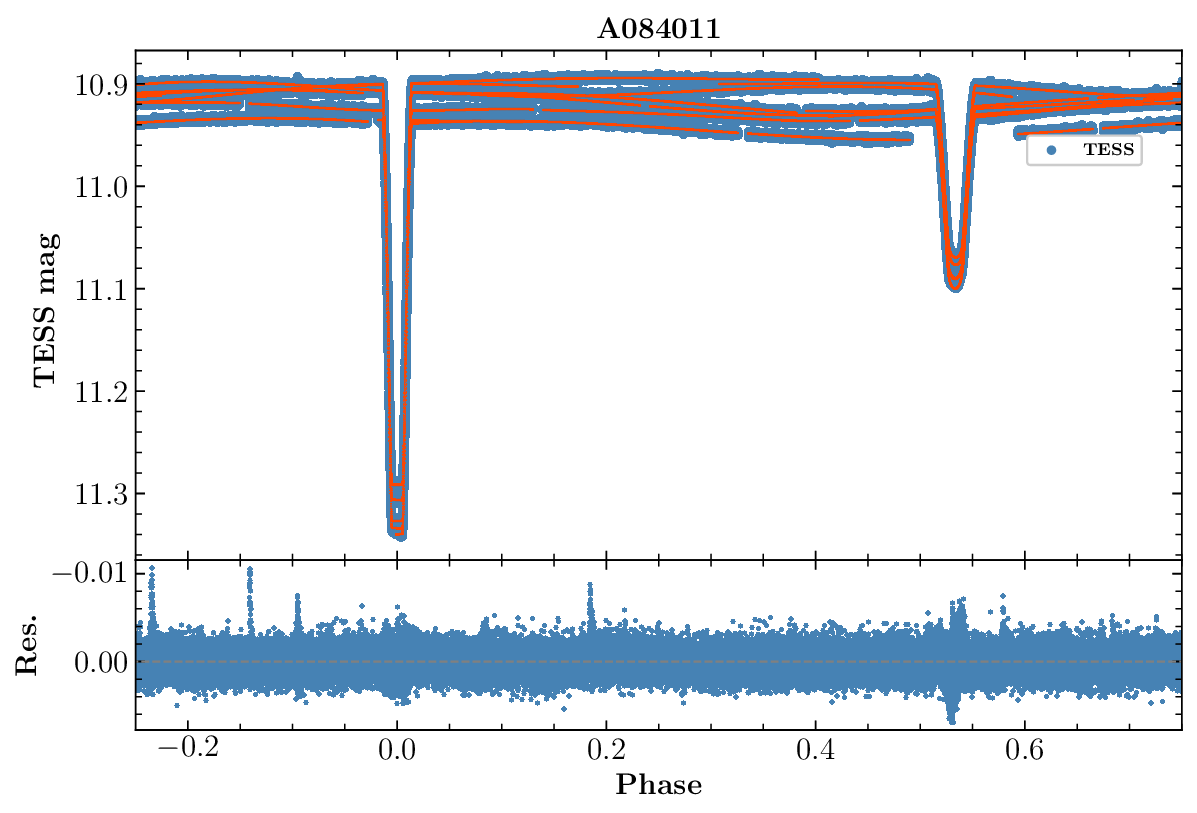}  %
    \includegraphics[width=0.33\textwidth]{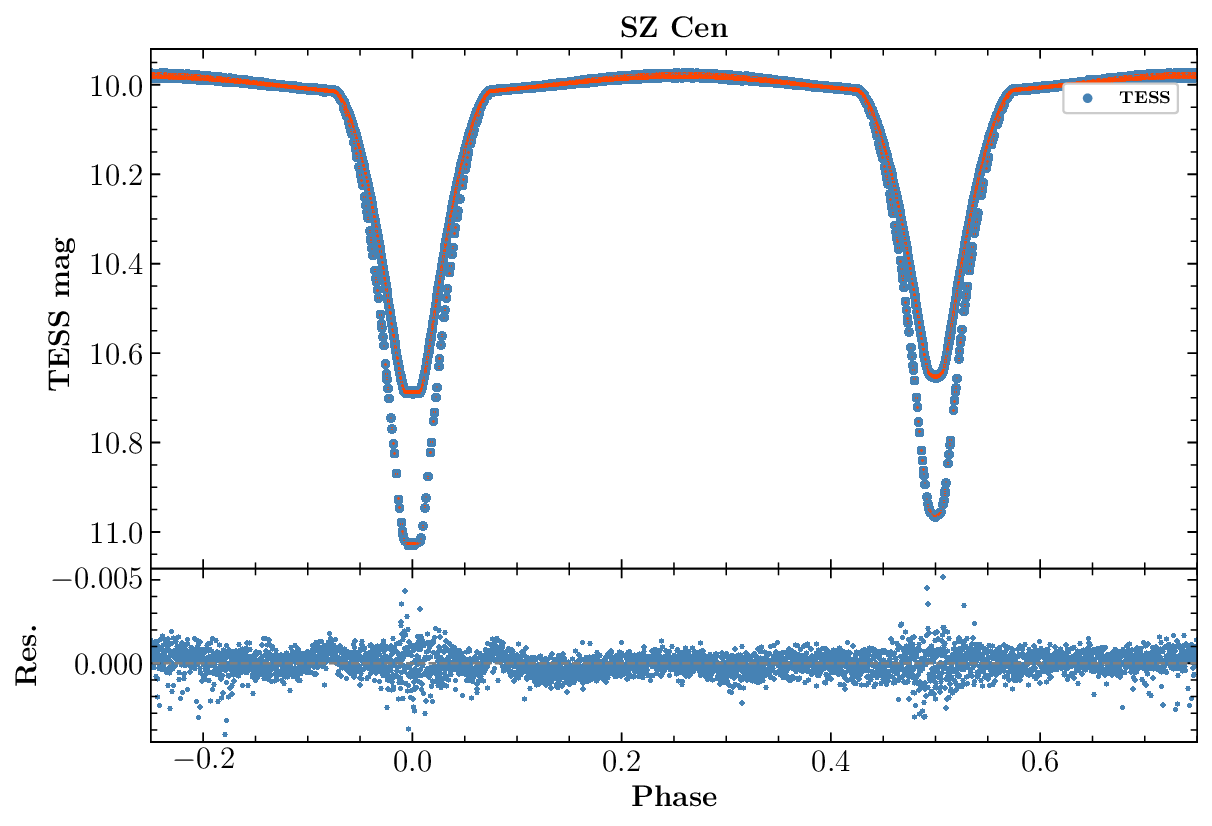}   %
    \includegraphics[width=0.33\textwidth]{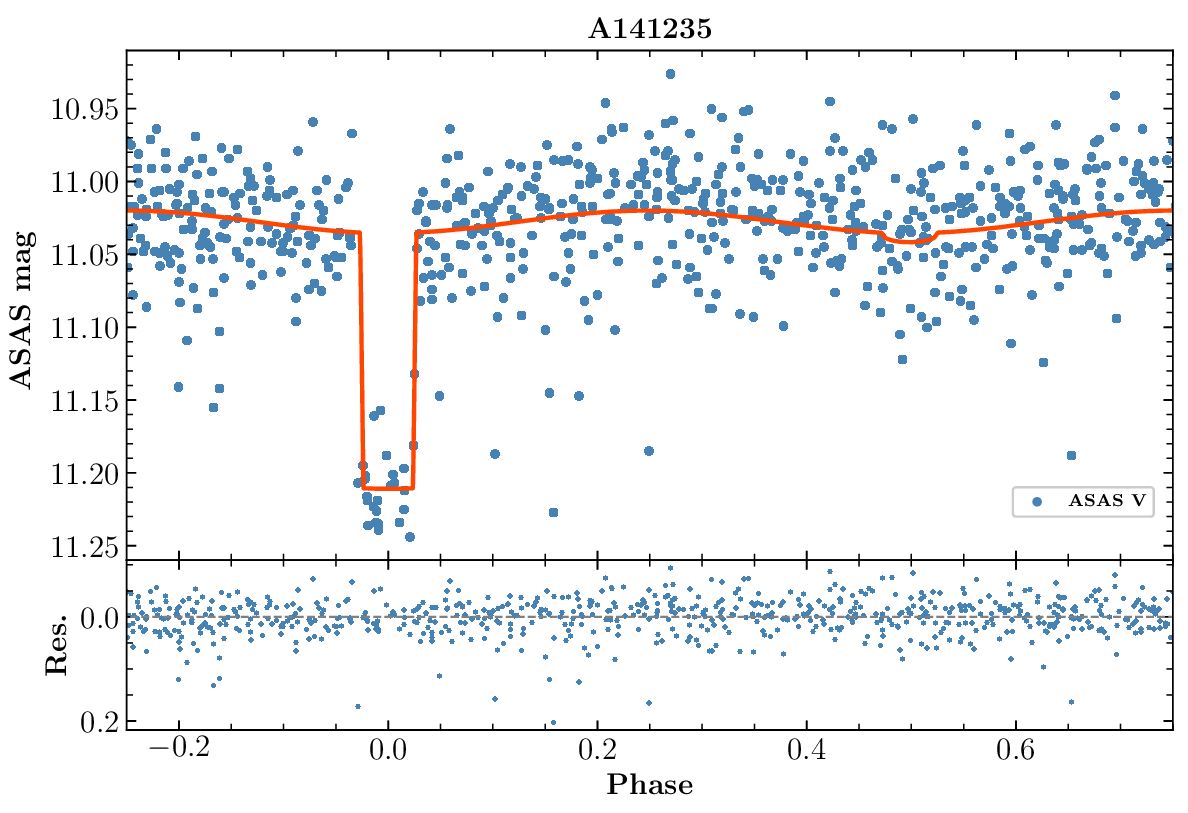}  %
    \includegraphics[width=0.33\textwidth]{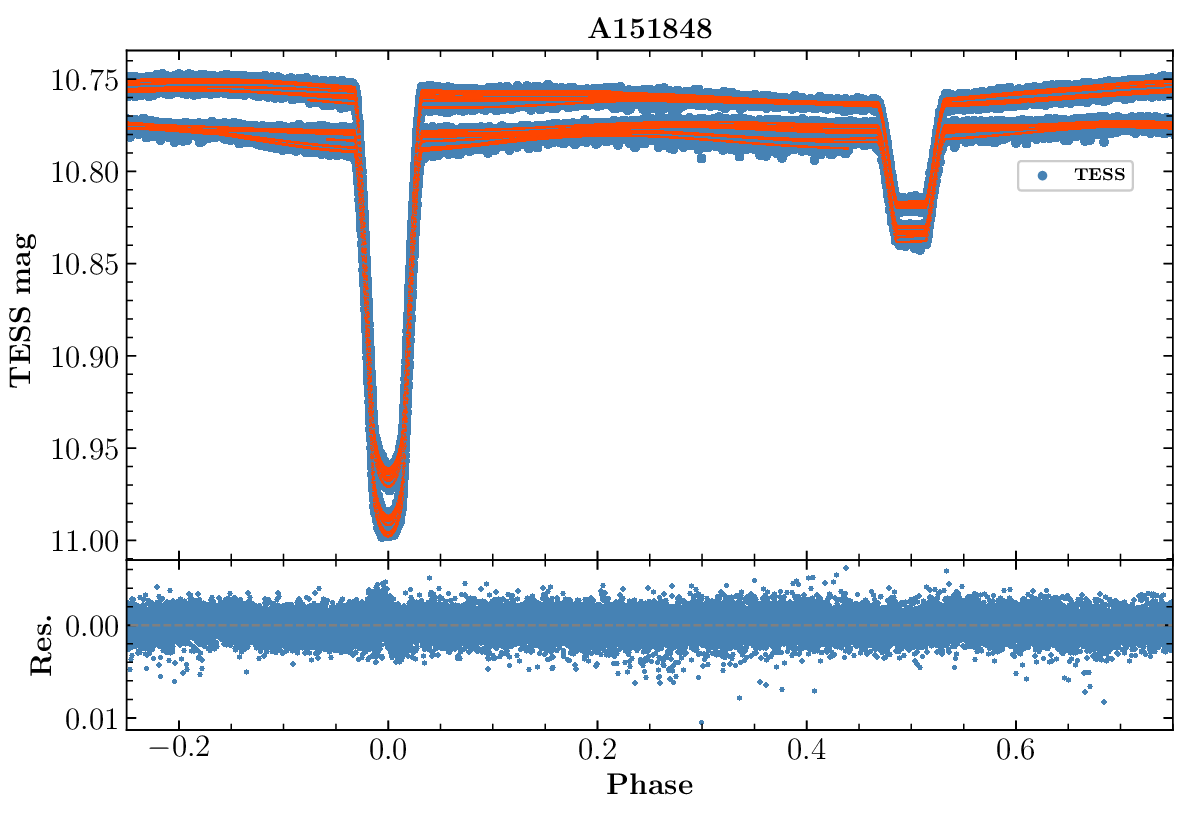}  %
    \includegraphics[width=0.33\textwidth]{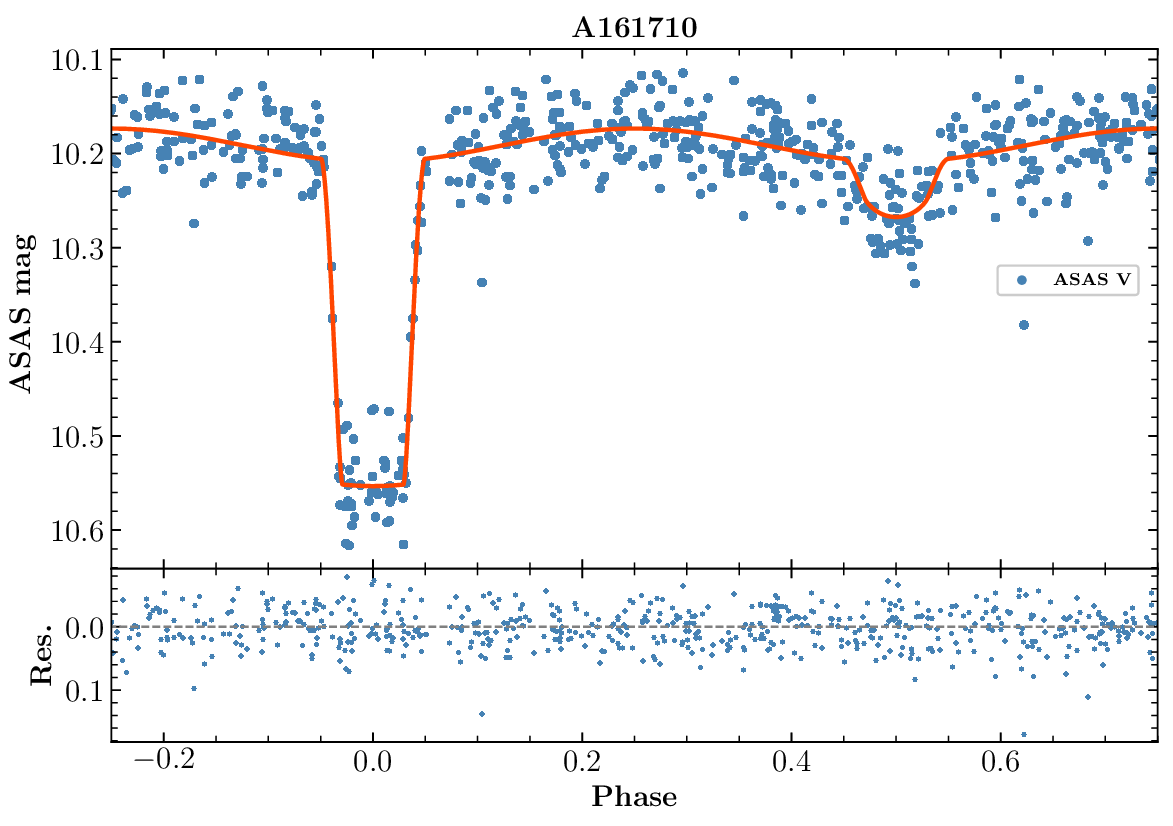}  %
    \includegraphics[width=0.33\textwidth]{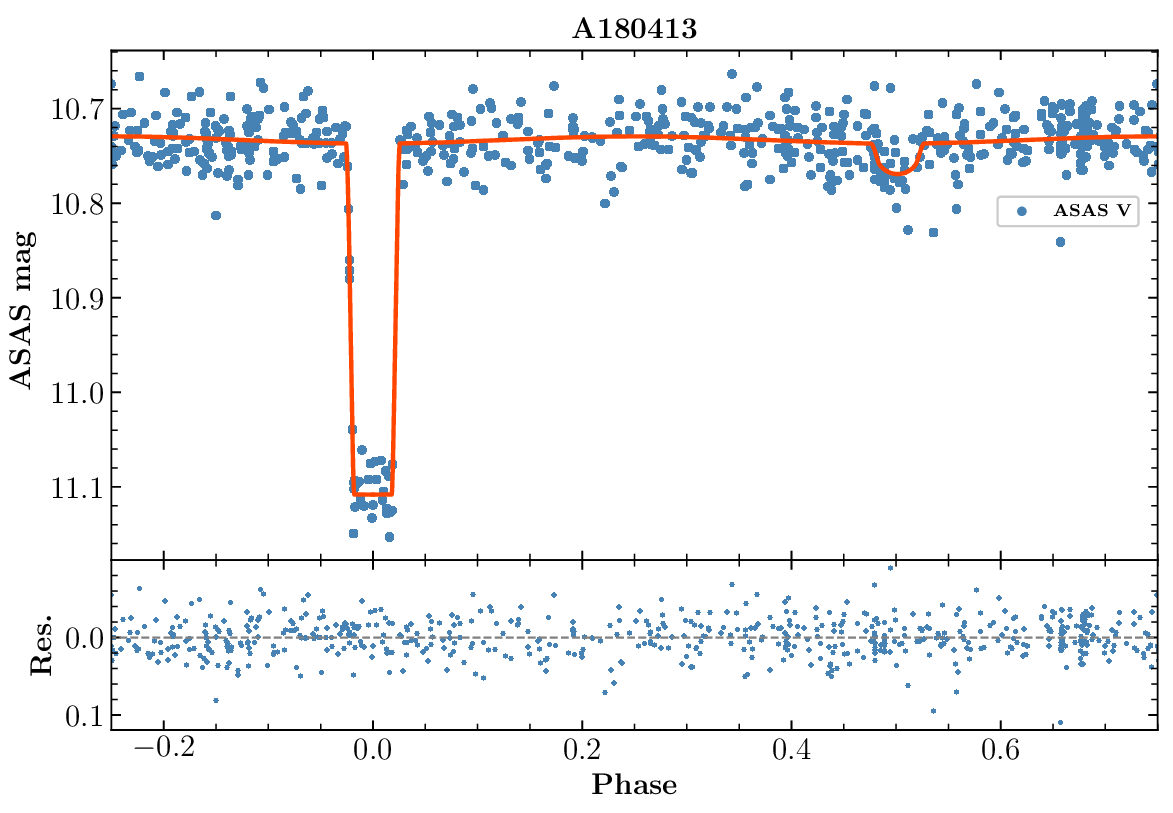}  %
    \includegraphics[width=0.33\textwidth]{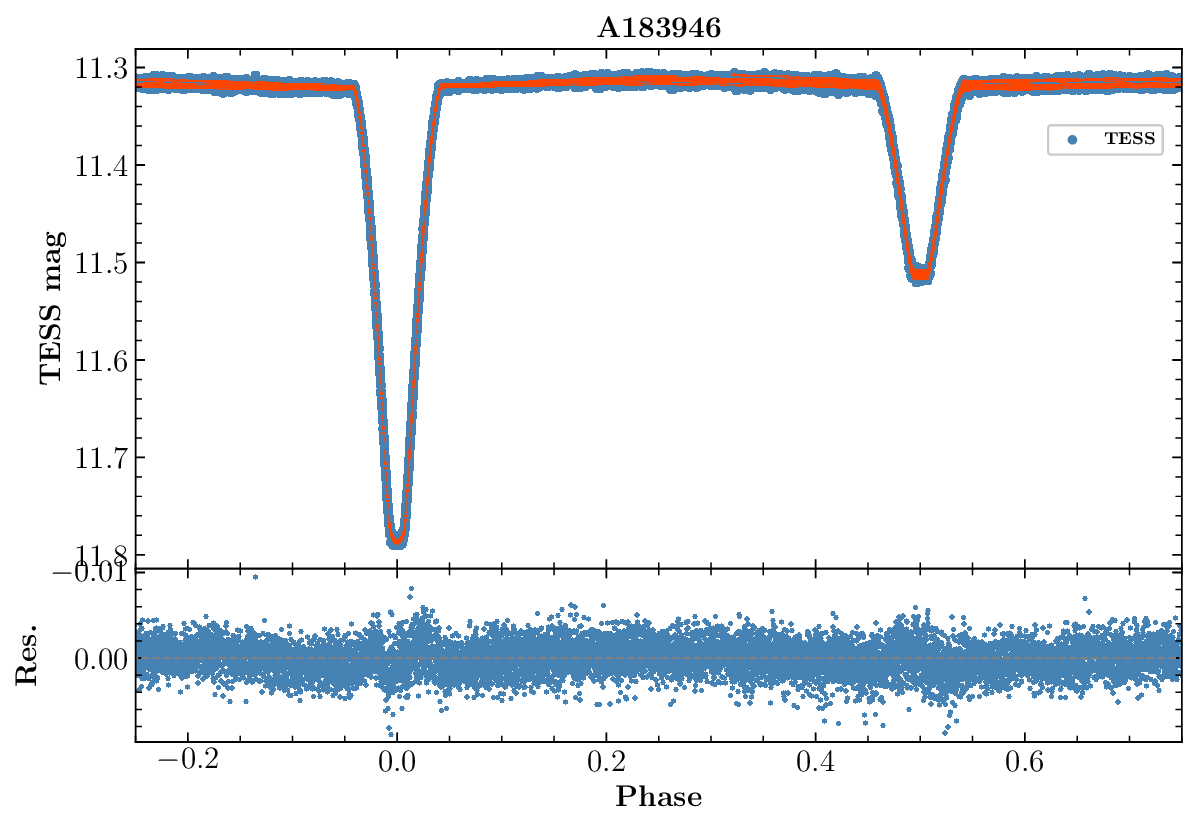}  %
    \includegraphics[width=0.33\textwidth]{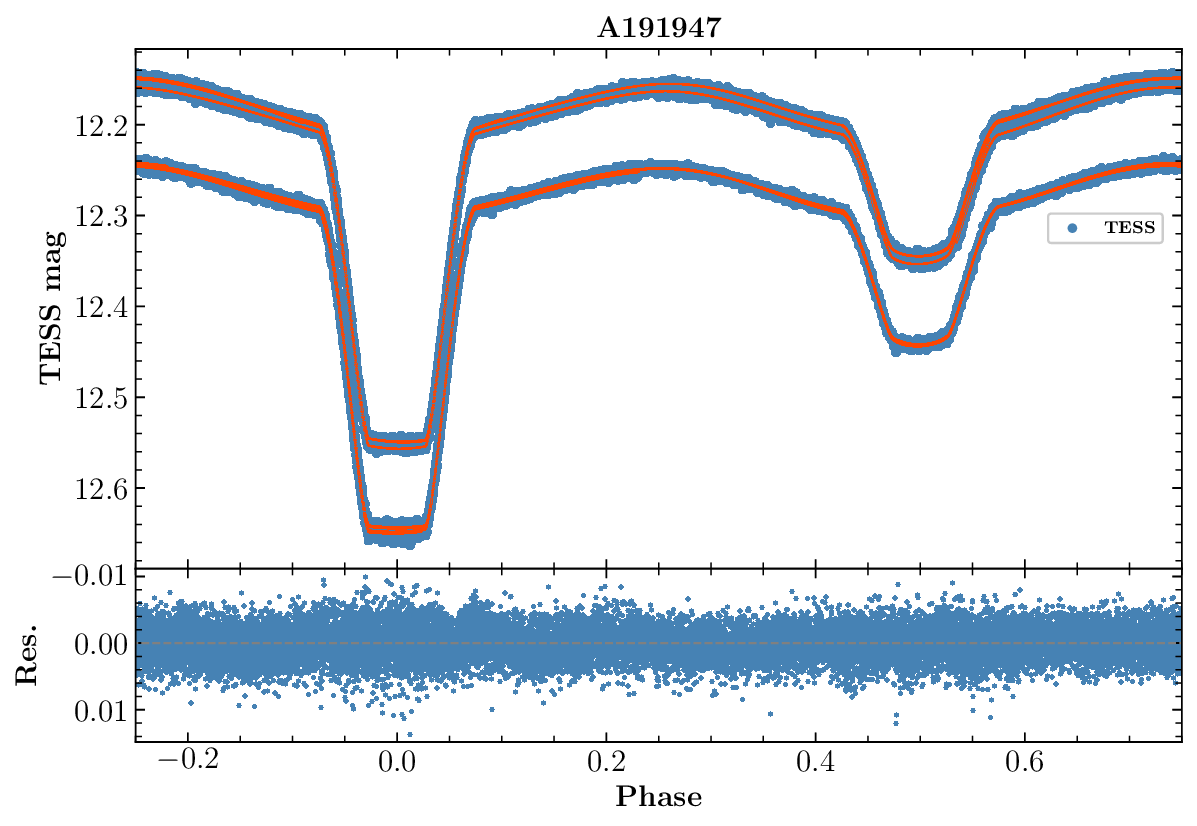}  %
    \includegraphics[width=0.33\textwidth]{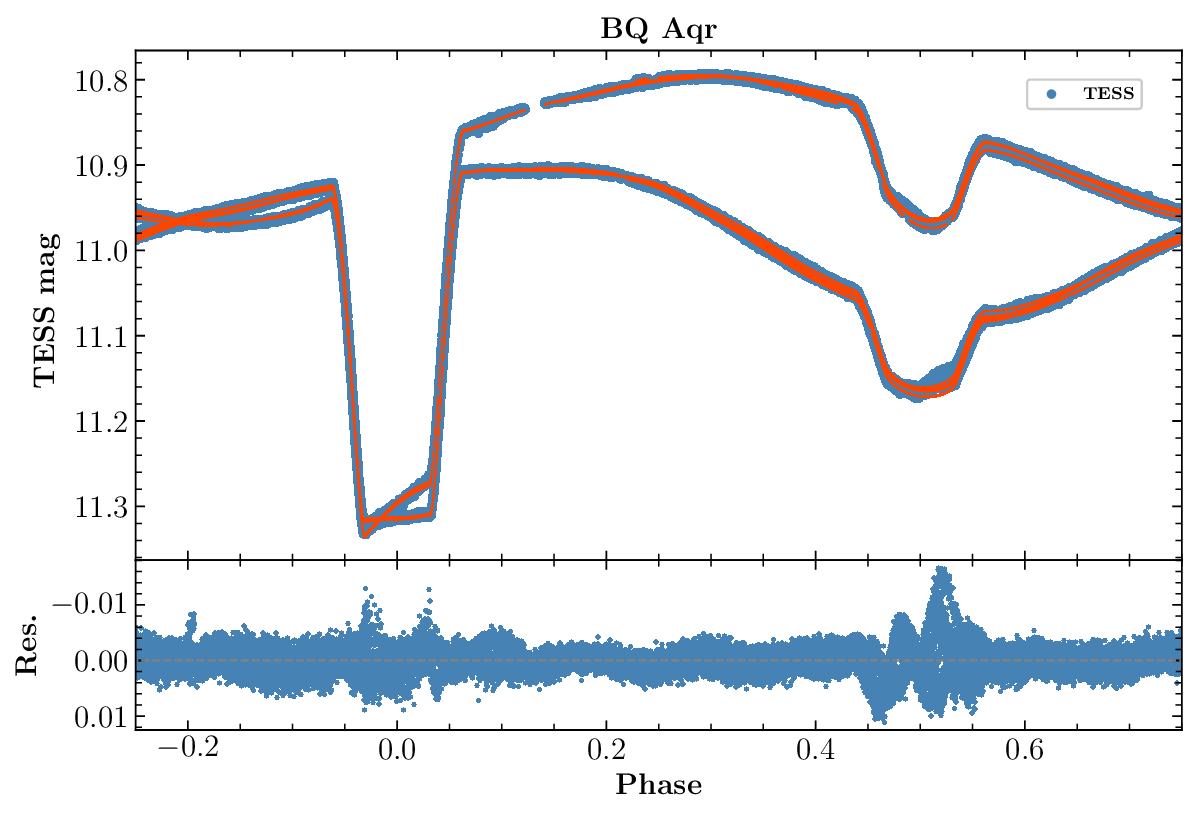}    %
    \caption{New and updated light curve solutions, based on TESS or ASAS photometry. Phase 0 is set to $T_0$ -- the mid-point of the deeper eclipse. The base level of the SAP flux may change between sectors (e.g. AL~Ari, A151848). For SZ~Cen, the nonphysically deep eclipses (S11, 1800~s cadence) are caused by overestimated level of third light contamination used in the QLP processing. Please note a different range of phases on the RZ~Eri panel.}
    \label{fig_lcs}
\end{figure*}

Here we show the model RV and LC curves, as well as the key binary parameters. Figures~\ref{fig_rvs} and \ref{fig_lcs} show the RV and light curves, respectively. Table~\ref{tab_paramsRVLC} summarises the orbital and physical parameters. Our results for four systems studied beforehand -- AL~Ari, RZ~Eri, SZ~Cen, and BQ~Aqr -- are compared with the literature in Tab.~\ref{tab_comparison}.

For most of the systems these results were obtained for the first time. We note a very good precision ($<$2\%) in both mass and radius estimates in the following cases: AL~Ari, SZ~Cen, A151848, and A183946. Two more objects (A191947, BQ~Aqr) have the same parameters known to $<$4\% level. Since some of the stars are in a short-lasting transition phase between the MS and red giant branch, they can be considered as valuable targets for studying this stage of evolution. One of them, RZ~Eri, probably underwent a substantial mass loss from at least one component, mass exchange episode(s), or a merger. The history of RZ~Eri is an interesting one, and more would be revealed with analysis of high-resolution disentangled spectra of both components, especially the MS one. 

A183946 was found to harbour a $\delta$~Scuti-type pulsator, whose major pulsation frequency was determined to be $f_{\rm puls}=9.4255$~d${-1}$ ($P_{\rm puls} = 0.106095$~d) with the amplitude of 0.663~mmag (TESS band, not corrected for the companion's flux). Other frequencies were also noted, but were much weaker, and their analysis is out of the scope of this paper. Another potentially interesting targets are A151848 with a low-mass (0.72~M$_\odot$) secondary, and A084011 residing in the PLATO's LOPS2 field.

\include{tab_params}

\section{Test of the light curve fitting approach.}




Despite \jktebop\ is a widely used light curve fitting code, it is still a simple geometric approach with some limitations, especially when it comes to short-period eclipsing binaries with prominent ellipsoidal variations. The typical ``rule of thumb'' limit for application of \jktebop\ is that the oblateness of the star should not exceed 4\% \citep{popp81}, and the fractional radius should be smaller than 0.25, above which the resulting radius may be wrong by more than 1\% \citep{north04}.
This is because the shapes of stars in \jktebop\ are approximated by spheres for calculating eclipse shapes, and as bi-axial ellipsoids for calculating proximity effects, while in more sophisticated codes, like the \texttt{WD} \citep{wd} or the PHysics Of Eclipsing BinariEs \citep[\texttt{PHOEBE};][]{prsa_phoebe,conroy_phoebe} it is dictated by the Roche geometry. Since our cases reach and slightly exceed the boundaries cited above, we decided to verify our results, especially how much our radii may be affected by the use of simplified stellar shapes.

We would like to stress that no general conclusions about the use of \jktebop\ should be made on the basis of our tests. These are meant only to verify our most extreme examples.

The approach to model a light curve of a DEB with stronger ellipsoidal variations, by applying \jktebop\ with additional sines and polynomials, has been tested with synthetic data. We used the latest version (v2.4) of \texttt{PHOEBE} to generate light curves of DEBs with substantial ellipsoidal variations. We based our synthetic binaries on three of our systems with TESS photometry available, and largest oblatenesses and fractional radii in the sample (given respectively in parentheses): SZ~Cen (2.3\%, 0.251), A191947 (4.4\%, 0.326), and BQ~Aqr (3.4\%, 0.290). We added a DEB based on the well-studied AL~Ari to ensure the procedure works correctly. We provided the code with input stellar parameters (i.e. $M, R, T_{\rm eff}, P, T_0$) from Tabs.~\ref{tab_paramsRVLC} and \ref{tab_params_ispec}, and used exactly the same times of observations as in real TESS curves. Finally, we added white noise corresponding to the $rms$ of the LC fit.

\begin{figure}
    \centering
    \includegraphics[width=0.9\columnwidth]{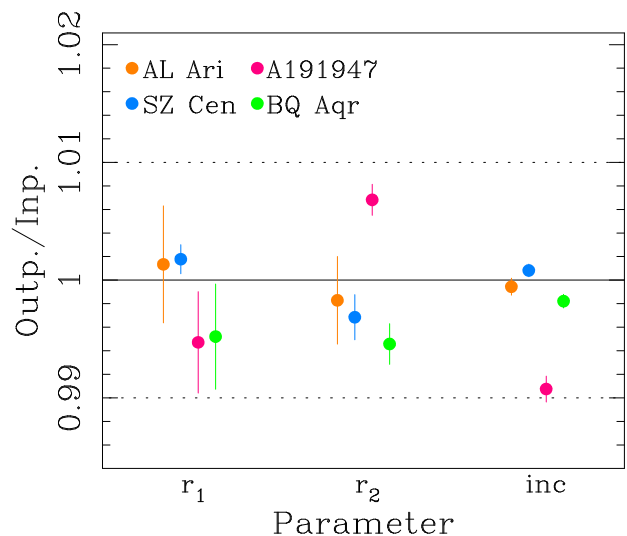}
    \caption{Comparison of results of the \jktebop\ analysis of synthetic light curves with the input values, for a sample of crucial parameters (fractional radii and inclination). The plot shows ratio of the output to the input value, with scaled uncertainties calculated with Monte-Carlo approach (Task~8). Each colour codes the analysis of a synthetic curve based on one of the selected systems. The dashed vertical lines mark $\pm$1\% deviation from the input. In every case, including those with the largest oblateness, \jktebop\ reproduced the input values with accuracy better than 0.7\%.}
    \label{fig_phoebe}
\end{figure}

These synthetic curves were analysed with \jktebop\ in the same way as described
in Sect. 4.1, including sine functions and polynomials, and errors estimated with a Monte Carlo procedure (Task~8). The MC errors may themselves be lower than the true precision, since they refer to single-sector cases, and do not include scatter between results obtained for different TESS sectors. The resulted values of inclination ($i$) and fractional radii ($r_1$ and $r_2$) are compared to their input values in Fig.~\ref{fig_phoebe}. One can see that \jktebop\ managed to reproduce input values of $r_{1,2}$ with accuracy better than 0.7\%, exceeding 0.5\% in only two cases (secondaries of A191947 and BQ~Aqr). Individual discrepancies are better than or comparable to the uncertainties quoted in Tab.~\ref{tab_paramsRVLC}. We thus conclude that \jktebop\ produced reliable results for the studied cases.

\end{appendix}
\end{document}

%% file: tab_params.tex
\begin{sidewaystable}
\centering
\scriptsize
\caption{Orbital and physical parameters of the studied systems. Values that were held fixed have ``fix'' instead of errors. The ``---'' symbol denotes parameters that are impossible to determine for a given system.}\label{tab_paramsRVLC}
\begin{tabular}{lccccccccccc}
\hline \hline
ID  & AL Ari & RZ Eri & A084011 & SZ Cen & A141235 & A151848 & A161710 & A180413 & A183946 & A191947 & BQ Aqr \\
\hline
$P$ (d) & 3.7474571(13) & 39.2838136(22) & 22.119211(70) & 4.1079819(43) & 160.703(76) & 3.6574068(11) & 10.132985(42) & 49.1110(9) & 2.1612188(19) & 4.713353(22) & 6.62032606(17) \\
$T_0$ (JD-2400000)$^a$ & 59450.899937(8) & 58462.13758(6) & 58564.661055(32) & 59334.524623(15) & 52011.66(11) & 56909.36019(12) & 51913.8792(59) & 51956.708(36) & 60481.359219(14) & 59772.09828(7) & 58360.261012(4) \\
$T_P$ (JD-2400000)$^b$ & 55466.016(46) & 58451.57679(6) & 55976.70(21) & 41385.72(12) & 55960(11) & 56122.0086(14) & 56349.49(26) & 55963.6(3.2) & 56785.1386(9) & 56787.382(45) & 56101.0284(24) \\
$K_1$ (km/s) & 76.949(30) & 47.9(1.5) & 56.31(79) & 111.33(62) & --- & 65.01(35) & 81.56(44) & --- & 94.67(18) & 110.2(1.6) & 85.18(13) \\
$K_2$ (km/s) & 98.204(80) & 52.0(4) & 52.4(1.3) & 109.41(61) & 31.91(8) & 105.99(65) & 66.1(1.4) & 45.35(18) & 133.06(42) & 97.1(8) & 79.3(1.4) \\
$\gamma_1$ (km/s) & -19.027(69) & 46.3(9) & 6.48(81) & -20.05(33) & -24.74(6) & 92.08(22) & -20.86(49) & -0.42(11) & -8.75(16) & -37.15(26) & 19.09(30) \\
$\gamma_2-\gamma_1$ (km/s) & 0.0(fix) & 0.0(fix) & 0.0(fix) & 0.0(fix) & 0.0(fix) & 0.0(fix) & 1.8(1.3) & 0.0(fix) & 0.7(1.2) & 0.77(76) & 0.6(2.7) \\
$q$ & 0.7836(8) & 0.921(30) & 1.074(31) & 1.017(8) & --- & 0.6134(50) & 1.234(26) & --- & 0.7115(26) & 1.135(19) & 1.073(19) \\
$M_1\sin^3(i)$ (M$_\odot$) & 1.1641(22) & 1.688(59) & 1.359(70) & 2.269(28) & --- & 1.174(17) & 1.51(6) & --- & 1.545(11) & 2.038(45) & 1.473(52) \\
$M_2\sin^3(i)$ (M$_\odot$) & 0.9121(13) & 1.554(98) & 1.460(55) & 2.309(29) & --- & 0.720(9) & 1.87(4) & --- & 1.099(6) & 2.313(72) & 1.582(28) \\
$a\sin(i)$ (R$_\odot$) & 12.956(7) & 72.0(1.1) & 46.85(66) & 17.93(7) & --- & 12.365(53) & 29.58(29) & --- & 9.731(20) & 19.32(17) & 21.54(19) \\
$e$ & 0.0572(4) & 0.3721(16) & 0.169(4) & 0.0(fix) & 0.0052(21) & 0.0(fix) & 0.0(fix) & 0.003(1) & 0.0(fix) & 0.0(fix) & 0.0(fix) \\
$\omega$($^\circ$) & 72.8(1.6) & 315.2(5) & 72.0(4) & --- & 119(24) & --- & --- & 317(32) & --- & --- & --- \\
$r_1$ & 0.10493(64) & 0.03916(10) & 0.03121(17) & 0.20266(16) & 0.0116(5) & 0.14414(46) & 0.0619(43) & 0.0227(18) & 0.15270(89) & 0.1340(31) & 0.09172(41) \\
$r_2$ & 0.06991(18) & 0.09660(14) & 0.07280(42) & 0.25060(46) & 0.163(10) & 0.05840(12) & 0.2436(21) & 0.1428(80) & 0.10774(24) & 0.3255(28) & 0.28960(49) \\
$i$ ($^\circ$) & 89.42(11) & 88.86(2) & 88.77(7) & 88.75(28) & 87.6(2.3) & 89.46(74) & 89.9(1) & 87.4(1.4) & 89.09(6) & 84.6(6) & 89.5(5) \\
$J^c$ & 0.635(27)T & 0.2672(68)T & 0.423(14)T & 0.906(15)T & 0.029(21)A & 0.340(11)T & 0.158(12)A & 0.071(12)A & 0.514(16)T & 0.384(14)T & 0.226(26)T \\
${l_2/l_1}^c$ & 0.2697(65)T & 1.48291(12)T & 2.251(45)T & 1.3950(52)T & 5.80(33)A & 0.0524(9)T & 2.38(25)A & 2.537(38)A & 0.2473(13)T & 2.60(13)T & 2.32(12)T \\
${l_3/l_{\rm tot}}^{c,d}$ & 0.0(fix) & 0.0(fix) & (var) & (var) & 0.0(fix) & 0.017(19)T & 0.111(58)A & 0.0(fix) & 0.176(4)T & 0.077(28)T & 0.0(fix) \\
$rms_{\rm RV1}$ (km/s) & 0.19 & 3.80 & 0.49 & 2.18 & --- & 0.17 & 1.22 & --- & 0.25 & 0.76 & 0.43 \\
$rms_{\rm RV2}$ (km/s) & 0.34 & 2.10 & 1.35 & 1.73 & 0.09 & 0.84 & 3.98 & 0.18 & 2.14 & 1.79 & 4.79 \\
$rms_{\rm LC}$ (mmag) & 0.71 & 0.86 & 1.23 & 0.69 & 36 & 1.15 & 30.67 & 26.30 & 1.75 & 2.43 & 2.43 \\
$M_1$ (M$_\odot$) & 1.1643(22) & 1.689(59) & 1.359(69) & 2.271(28) & --- & 1.175(17) & 1.51(6) & --- & 1.546(11) & 2.065(46) & 1.474(52) \\
$M_2$ (M$_\odot$) & 0.9123(13) & 1.555(98) & 1.460(54) & 2.311(29) & --- & 0.721(9) & 1.87(4) & --- & 1.100(6) & 2.344(73) & 1.582(28) \\
$a$ (R$_\odot$) & 12.957(7) & 72.0(1.1) & 46.87(66) & 17.93(7) & --- & 12.366(53) & 29.58(29) & --- & 9.732(20) & 19.40(17) & 21.54(19) \\
$R_1$ (R$_\odot$) & 1.3596(83) & 2.820(45) & 1.463(22) & 3.634(15) & --- & 1.7825(96) & 1.83(13) & --- & 1.4861(92) & 2.600(65) & 1.975(91) \\
$R_2$ (R$_\odot$) & 0.9058(24) & 6.96(11)  & 3.412(52) & 4.494(20) & --- & 0.7222(34) & 7.21(9)  & --- & 1.0485(31) & 6.317(78) & 6.237(55) \\
$\log(g_1)$ & 4.2376(53) & 3.765(4) & 4.246(13) & 3.674(3) & --- & 4.006(4) & 4.093(61) & --- & 4.283(5) & 3.923(21) & 4.015(40) \\
$\log(g_2)$ & 4.4844(22) & 2.945(14) & 3.538(6) & 3.497(3) & --- & 4.579(3) & 2.994(8)  & --- & 4.438(2) & 3.207(10) & 3.0475(16) \\
\hline 
\end{tabular}\\
$^a$ Mid-time of the primary (deeper) eclipse. $^b$ Time of pericentre or quadrature. 
$^c$ ``T'': in TESS filter, ``A'': in ASAS V filter.\\
$^d$  The value in TESS band can vary from sector to sector, depending on the orientation of the satellite or the amount used in the detrending process.\\
\end{sidewaystable}

\begin{sidewaystable}
\centering
\scriptsize
\caption{Comparison of our results with the literature. This work's values of ages, and companion's temperature and metallicity are taken from distribution of the accepted models. Same values for the spectroscopic component come from \ispec\ (Tab.~\ref{tab_params_ispec}). Note that the given [$M/H$] is current, not initial like in Tab.~\ref{tab_iso}. Literature values of $T_{\rm eff}$ for RZ~Eri and SZ~Cen are based on colour indices, not spectroscopic analysis. Values assumed by the authors are given with no error, i.e. with ``-'' in parenthesis.}
\label{tab_comparison}
\begin{tabular}{l|cc|ccc|ccc|cc}
\hline \hline
Target  & \multicolumn{2}{c|}{AL Ari} & \multicolumn{3}{c|}{RZ Eri} & \multicolumn{3}{c|}{SZ Cen} & \multicolumn{2}{c}{BQ Aqr} \\
\hline
Param.  &   This work   & \citet{grac21}  & This work   & \citet{popp88} & \citet{vive} & This Work & \citet{ande75} & \citet{gron77} & This work & \citet{rata16} \\
\hline
$P$ [d]             & 3.7474571(13)$^a$& 3.7474513(8)$^a$& 39.2838136(22) & 39.28238(-) & 39.282466(-) & 4.1079819(43)	& 4.107983(-) &	4.107983(-) & 6.2032606(17) & 6.6205062(21)	\\		
$i$ [$^\circ$]      & 89.42(11)  & 89.30(8) 		& 88.86(2)  & 89.0(1.0) & 89.61(7)	& 88.75(28) & 88.4(3)   & 88.03(30)    & 89.5(5)   & 89.5(5)  	\\
$M_1$ [M$_\odot$]   & 1.1643(22) & 1.1640(13)	 	& 1.69(6)   & 1.68(10)  & 1.69(6)	& 2.271(28) & 2.276(21) & 2.277(21)    & 1.474(52) & 1.490(40)	\\	
$M_2$ [M$_\odot$]   & 0.9123(13) & 0.9112(7)		& 1.55(10)  & 1.62(20)  & 1.63(13)	& 2.311(29) & 2.316(26) & 2.317(26)    & 1.582(28) & 1.588(21)	\\	
$R_1$ [R$_\odot$]   & 1.3596(83) & 1.3720(90) 	 	& 2.82(5)   & 2.83(-)   & 2.84(21) 	& 3.634(15) & 3.620(21) & 3.620(21)    & 1.975(91) & 2.072(14)	\\	
$R_2$ [R$_\odot$]   & 0.9058(24) & 0.9050(80)		& 6.96(11)  & 7.00(30)  & 6.94(20)	& 4.494(20) & 4.534(23) & 4.552(23)    & 6.237(55) & 6.53(28) 	\\	
$T_{\rm eff,1}$ [K] & 6540(100)  & 6578(82) & 7600($^{+130}_{-790}$) & 6470(570)  & 7400(-)   & 8506($^{+96}_{-289}$) & 8130(-)   & 8280(300) & 7400($^{+67}_{-380}$)& 6390(230)\\
$T_{\rm eff,2}$ [K] & 5690($^{+120}_{-14}$) & 5560(192)	& 4890(170)  & 4520(230)  & 4670(70)  & 7689(94)	& 7760(-) & 8000(300) & 4630(110) & 4490(230)	\\ 	
$[M/H]_1$ [dex]     & -0.38(5)  & -0.42(7)  & -0.30(15) & --- & --- & 0.00($^{+0.07}_{-0.05}$) & 0.0(-) & 0.0(-) & -0.24($^{+0.12}_{-0.05}$)& 0.12(11)\\
$[M/H]_2$ [dex]     & -0.19(5)  & -0.60(13)	& -0.29(18) & --- &	--- & 0.00(5) &	0.0(-) & 0.0(-) & -0.25(9)  & 0.12(11) 	\\
$\tau$ [Gyr]        & 3.24($^{+0.08}_{-0.35}$) & --- & 1.38/2.04$^b$ & --- &  2.5(-) & 0.724($^{+0.052}_{-0.024}$) & 0.5(-) & 0.45(-) & 2.09($^{+0.31}_{-0.09}$) & 2.50/2.65$^c$ \\
\hline
\end{tabular}\\
$^a$ Values in agreement when $\dot P$ from \citet{grac21} is taken into account.
$^b$ For each component separately, under two different evolutionary scenarios. \\
$^c$ No age uncertainties are given by \citet{rata16}. Instead two separate solutions are presented.
\end{sidewaystable}